\documentclass[oldversion]{aa}
\usepackage{graphicx}
\usepackage{txfonts}
\usepackage{natbib}
\usepackage{longtable}

\begin{document}
   \title{The Brera Multi-scale Wavelet \mbox{\boldmath Chandra} 
	Survey\thanks{The catalogue is available in electronic form at the CDS 
	via web {\tt  http://cdsweb.u-strasbg.fr/cgi-bin/qcat?J/A+A/} 
	or via anonymous ftp to {\tt cdsarc.u-strasbg.fr (130.79.128.5)}.}}
   \subtitle{I. Serendipitous source catalogue}
   \author{P.\ Romano\inst{1,2}  
	\and S.\ Campana\inst{1}  
	\and R.P.\ Mignani\inst{3} 
	\and A.\ Moretti\inst{1} 
	\and M.\ Mottini\inst{4} 
	\and M.R.\ Panzera\inst{1} 
	\and G.\ Tagliaferri\inst{1}
	}
   \offprints{P.\ Romano; \email{romano@ifc.inaf.it} }
   \institute{INAF--Osservatorio Astronomico di Brera,
              Via E.\ Bianchi 46, 23807 Merate (LC), Italy 
	\and INAF, Istituto di Astrofisica Spaziale e Fisica Cosmica, 
              Via U.\ La Malfa 153, I-90146 Palermo, Italy 
	\and  Mullard Space Science Laboratory, University College London, 
	      Holmbury St. Mary, Dorking, Surrey, RH5 6NT, United Kingdom
	\and  European Southern Observatory,
                Schwarzschild Stra\ss e  2,
                85740 Garching bei M\"unchen, Germany
	}
   \date{Received ; accepted }


\abstract

We present the BMW-{\it Chandra}  source catalogue 
drawn from essentially all {\it Chandra}   ACIS-I pointed observations with an
exposure time in excess of 10\,ks public as of March 2003 (136 observations).
Using the wavelet detection algorithm developed by Lazzati et al.\ (1999) and
Campana et al.\ (1999), which can characterise both point-like and extended
sources, we identified 21325 sources. 
Among them, 16758 are serendipitous, i.e.\ not associated with the targets 
of the pointings, and do not require a non-automated analysis. 
This makes our catalogue the largest compilation of {\it Chandra} sources to date. 
The 0.5--10\,keV absorption corrected fluxes of these sources 
range from $\sim 3\times 10^{-16}$ to $9\times10^{-12}$ erg cm$^{-2}$ s$^{-1}$ with a median of 
$7\times 10^{-15}$ erg cm$^{-2}$ s$^{-1}$.
The catalogue consists of count rates and relative errors in three energy bands
(total, 0.5--7\,keV; soft, 0.5--2\,keV; and hard, 2--7\,keV), and 
source positions relative to the highest signal-to-noise detection among the three bands. 
The wavelet algorithm also provides an estimate of the extension of 
the source.
We include information drawn from the headers of the original files, as well, and
extracted source counts in four additional energy bands, 
SB1 (0.5--1\,keV), SB2 (1--2\,keV), HB1 (2--4\,keV), and HB2 (4--7\,keV).
We computed the sky coverage for the full catalogue and for a subset
at high Galactic latitude ($\mid b \mid \, > 20^{\circ}$).
The complete catalogue provides a sky coverage in the soft band (0.5--2\,keV, S/N =3)
of $\sim 8$ deg$^2$ at a limiting flux of $\sim 10^{-13}$  erg cm$^{-2}$ s$^{-1}$,
and $\sim 2$ deg$^2$ at a limiting flux of $\sim 10^{-15}$ erg cm$^{-2}$ s$^{-1}$.
Furthermore, we present the results of the cross-match with existing catalogues
at different wavelengths (FIRST, IRAS, 2MASS, GSC2, and ChaMP). 
The total numbers of matches with the FIRST, IRASPSC, 2MASS, and GSC2 catalogues 
obtained after a closest-distance selection are 13, 87, 6700, and 4485, respectively.

\authorrunning {P.\ Romano et al.}
\titlerunning {The Brera Multi-scale Wavelet {\it Chandra} 	Survey}
\maketitle


	\section{Introduction}

The use of a wavelet transform (WT) as an X-ray detection 
algorithm was pioneered by Rosati et al.\ (1995; 1998)
for the detection of extended sources in the Roentgen Satellite ({\it ROSAT\/}) 
Position Sensitive Proportional Counter (PSPC) fields 
and subsequently adopted by many groups 
(\citealt{Grebenevea95,Damianiea97b,Pislarea97,Vikhlininea98,Lazzatiea98,Freemanea02}). 
Differently from other WT-based algorithms, the Brera Multi-scale Wavelet 
(BMW, \citealt{Lazzatiea99,Campanaea99}) algorithm, which was developed to analyse 
{\it ROSAT} High Resolution Imager (HRI) data, automatically characterises 
each source through a multi-source $\chi^2$ fitting with respect to a Gaussian model 
in the wavelet space. 
For these reasons it has also proven to perform well in 
crowded fields and in conditions of very low background (\citealt{Lazzatiea99}).

	The BMW was used to produce the BMW-HRI catalogue (\citealt{Lazzatiea99,Campanaea99,Panzeraea03}), 
a catalogue of $\sim 29\,000$ sources detected with a probability of $\ga 4.2\sigma$ and with 
a sky coverage of 732 deg$^2$ down to a limiting flux of $10^{-12}$ erg cm$^{-2}$ s$^{-1}$
and 10 deg$^2$ down to a limiting flux of $10^{-14}$ erg cm$^{-2}$ s$^{-1}$. 

The BMW-HRI  is being  currently used for  a number of  scientific projects.
Among  preliminary  results,  an  analysis of  X-ray  detected  sources
without obvious counterparts at other wavelengths, (a.k.a. blank
field  sources, \citealt{Chieregatoea05}) has  been  carried out 
with  the   aim  of  identifying   unusual  objects,  as  well   as  a
serendipitous  X-ray  survey  of  Lyon-Meudon  Extragalactic  Database
(LEDA) normal galaxies (\citealt{Tajerea05}).

The BMW algorithm was modified to support the analysis of
{\it Chandra} Advanced CCD Imaging Spectrometer (ACIS) images 
(\citealt{Morettiea02}), and has led to interesting results on the nature of 
the cosmic X-ray background (CXB, \citealt{Campanaea01}; \citealt{Morettiea03}). 
Given the reliability and versatility of the BMW algorithm, we decided to apply it to a 
large sample of {\it Chandra} ACIS-I images, to take full advantage of the 
superb spatial resolution [$\sim 0\farcs5$ point-spread function (PSF) on-axis] 
of {\it Chandra} while being able to automatically analyse crowded fields 
and/or with very low background. We thus produced the 
Brera Multi-scale Wavelet {\it Chandra} source catalogue (BMW-C). 

In this  paper we present a pre-release of the BMW-C,  
which is based on a subset of the whole  {\it Chandra} ACIS observations dataset,
roughly  corresponding to  the first  three years  of  operations.  

Like the BMW-HRI, our catalogue provides source positions, count rates,
extensions and relative errors. In addition, for all bright  
(100 counts in the 0.5--7\,keV band) sources in the catalogue
we extracted light curves which will be exploited in a search for periodic  
and non-periodic variability, 
as well as spectra to have an immediate spectral classification.
 
In Sect.~\ref{bmwc:datasample} we describe the selection criteria of the 
{\it Chandra} fields and the resulting data sample.
In Sect.~\ref{bmwc:dataproc} we describe the data processing, 
which includes data screening and correction, event list selections 
and energy band selections. 
In Sect.~\ref{bmwc:detect} we describe the wavelet detection 
algorithm  and its application to the {\it Chandra} fields as well as the resulting source catalogue, 
the BMW-{\it Chandra} catalogue.
In Sect.~\ref{bmwc:character} we report the properties of the source sample, including
their spatial extent 
fluxes, and the definition of our serendipitous sub-sample.
In Sect.~\ref{bmwc:skycov_how} we calculate the sky coverage of our survey. 
In Sect.~\ref{bmwc:champcat} we compare our catalogue with the 
{\it Chandra} Multiwavelength Project (ChaMP, \citealt[][and references therein]{Champ3}) data.
In Sect.~\ref{bmwc:xcorr} we describe preliminary results of
the cross-matches between the 
BMW-{\it Chandra} and other catalogues at different wavelengths. 
Finally, in Sect.~\ref{bmwc:summary} we summarise our work and highlight the plan 
for future exploitation of the catalogue.

	\section{Data sample\label{bmwc:datasample}} 

	\begin{figure}	
	 	\resizebox{\hsize}{!}{\includegraphics{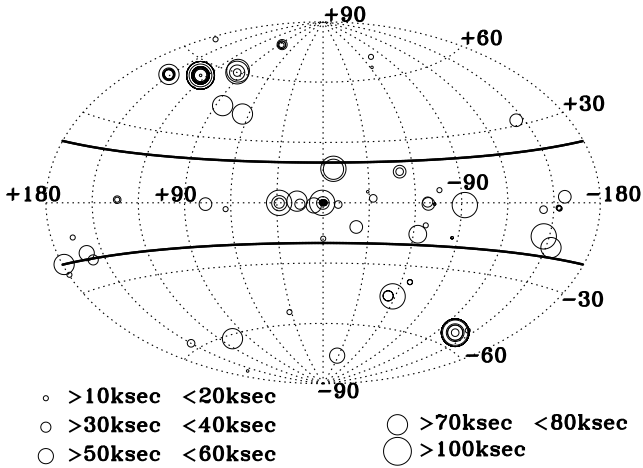}}
 		\caption{Aitoff projection in Galactic coordinates of the 
		136 observations in the BMW-C catalogue. 
		The size of each point is proportional to the exposure time,
		and the scale shows representative exposure times.} 
		The thick lines	are the limits of the high- and low-latitude 
		sub-catalogues (see Sect.~\ref{bmwc:character}).
 		\label{bmwc:aitoff}
	\end{figure}

Our choice of the {\it Chandra} fields favoured the ones that would 
maximise the sky area not occupied by the pointed targets, that is the
fields where the original PI was interested in a single, possibly 
point-like object centred in the field. We adopted the following criteria:
\begin{enumerate}
\item 	All ACIS-I [no grating, no High Resolution Camera (HRC) fields 
	and in Timed Exposure mode] 
	fields with exposure time in excess of 10\,ks available by 2003 March  
	were considered.
	Data from all four front-illuminated (FI) CCDs (I0, I1, I2, I3) were used. 
\item 	We excluded fields dominated by extended sources   
	[covering $\ga 1/9$ of the field of view (FOV)]. 
\item 	We excluded planet observations  and supernova remnant observations.
\item 	We also excluded fields with bright point-like or high-surface 
        brightness extended sources. 
\item 	We put no limit on Galactic latitude, but  
	we selected sub-samples based on latitude at a later time. 
\end{enumerate}
The exclusion of bright point-like or high-surface brightness 
extended sources was dictated by the nature of our detection algorithm, 
which leads to an excessive number of spurious detections at the 
periphery of the bright source. 
This problem is common to most detection algorithms.
Therefore, each field was visually inspected to check for such effects;
where found, a conservatively large portion of the field was flagged
(see Sect.~\ref{bmwc:detect}). 
Of the 147 fields analysed, 11 ($\sim 7$\%) were discarded because of problems
at various stages of the pipeline execution. 
The field relative to observation ID 634, for instance, was discarded because
of the spacecraft wobbling during the exposure that caused the images  
of point sources to be elongated and to be detected as double. 
We eliminated observation ID 581 because the CCD I3 suffered from good time intervals  
inconsistent with the other CCDs, resulting in an anomalously high background  
(e.g., see \citealt{Morettiea02}). 
Furthermore, the observation IDs  316, 361, 950, 1302, 1872, 2269, and  2271 
were eliminated because of problems at the detection stage.
Finally, the fields relative to observation IDs 2365 and 1431 were discarded 
at the analysis stage because the asset of the spacecraft changed during the 
observation. 
As a result, 136 fields reported successful completion of the pipeline. 
Figure~\ref{bmwc:aitoff} shows the Aitoff projection in 
Galactic coordinates of their positions. Table~\ref{bmwc:fieldprops}
lists the basic properties (observation ID, Sequence number, coordinates 
of the target, etc.) of the fields included in our sample.
We note that several fields were observed more than once.
These fields were considered as different pointings, so that the number of 
distinct fields is 94 (see Sec.~\ref{bmwc:skycov_how}).


\setcounter{table}{0}
\begin{table*}
\begin{center}
\caption{Field Properties in the BMW-C Sample.}
\label{bmwc:fieldprops}
\begin{tabular}{rllcccccc}
\hline
\hline
\noalign{\smallskip}
ID & Sequence & Target Name & Nominal& Nominal Dec   & Observation Date & MJD & Effective  & Average  $N_{\rm H}$ \\
  ({\tt OBSID})    & ({\tt SEQ\_NUM})  & ({\tt OBJECT})&  RA ({\tt RA\_NOM}) & ({\tt DEC\_NOM}) & ({\tt DATE\_OBS}) & & Exp.\  & ({\tt NH\_WAVG)} \\ 
               &                 &             & (J2000)       & (J2000)        &      &            & ({\tt TEFF}, s)        & ($10^{20}$ cm$^{-2}$) \\ 
\noalign{\smallskip}
\hline
\noalign{\smallskip}
19	&200017	&CRACLOUDCORE	&285.44943	&$-$36.97264	&2000$-$10$-$07T17:01:59&   51824.7  & 19705.67&       7.75    \\
50	&290019	&ETACARINAE	&161.25481	&$-$59.68144	&1999$-$09$-$06T19:49:15&   51427.8  & 9143.05 &       1.36$\times 10^{2}$     \\
65	&290034	&NGC2516	&119.58374	&$-$60.78898	&1999$-$08$-$26T15:20:50&   51416.6  & 7754.19 &       1.46$\times 10^{1}$     \\
79	&300004	&NGC6397	&265.17007	&$-$53.68983	&2000$-$07$-$31T15:31:33&   51756.6  & 48343.20&       1.38$\times 10^{1}$     \\
88	&400001	&CENX$-$3	&170.31691	&$-$60.61832	&2000$-$11$-$25T07:57:31&   51873.3  & 35742.88&       1.19$\times 10^{2}$     \\
90	&400003	&4U1538$-$52	&235.60481	&$-$52.38390	&2000$-$04$-$08T22:46:16&   51642.9  & 20950.14&       9.71$\times 10^{1}$     \\
96	&400009	&GS2000+25	&300.70609	&25.23087	&1999$-$11$-$05T11:03:12&   51487.5  & 18836.82&       6.55$\times 10^{1}$     \\
322	&600082	&NGC4472	&187.44366	&8.00878	&2000$-$03$-$19T12:15:22&   51622.5  & 10363.09&       1.66    \\
324	&600084	&NGC4636	&190.71155	&2.69179	&1999$-$12$-$04T23:48:01&   51517.0  & 6296.65 &       1.81    \\
441	&900003	&AXAFSouthernDee&53.11186	&$-$27.80500	&2000$-$05$-$27T01:18:05&   51691.1  & 54985.14&       9.00$\times 10^{-1}$    \\
446	&200036	&W3B		&36.41589	&62.09362	&2000$-$04$-$03T02:57:18&   51637.1  & 20059.54&       9.07$\times 10^{1}$     \\
522	&800030	&A209		&23.01838	&$-$13.56364	&2000$-$09$-$09T09:50:41&   51796.4  & 9965.00 &       1.66    \\
580	&900001	&HUBBLEDEEPFIELD&189.31799	&62.21036	&1999$-$11$-$13T01:14:18&   51495.1  & 49433.22&       1.49    \\
582	&900003	&AXAFSouthernDee&53.11215	&$-$27.80473	&2000$-$06$-$03T02:38:23&   51698.1  & 128609.01&      9.00$\times 10^{-1}$    \\
606	&200031	&IC348		&56.12549	&32.13877	&2000$-$09$-$21T19:58:42&   51808.8  & 51663.77&       1.37$\times 10^{1}$     \\
611	&200036	&W3B		&36.41431	&62.09313	&2000$-$03$-$23T11:59:55&   51626.5  & 18467.16&       9.07$\times 10^{1}$     \\
633	&200058	&NGC3603	&168.78081	&$-$61.26516	&2000$-$05$-$01T23:29:33&   51666.0  & 46210.94&       1.35$\times 10^{2}$     \\
635	&200060	&RHOOPHCORE	&246.82513	&$-$24.57316	&2000$-$04$-$13T18:33:23&   51647.8  & 76286.52&       1.40$\times 10^{1}$     \\
637	&200062	&RHOOPHA	&246.64668	&$-$24.38723	&2000$-$05$-$15T23:35:16&   51680.0  & 96094.30&       1.41$\times 10^{1}$     \\
642	&200067	&NGC1333	&52.27541	&31.32702	&2000$-$07$-$12T22:55:58&   51738.0  & 36044.84&       1.51$\times 10^{1}$     \\
652	&300021	&V382Velorum1999&161.25778	&$-$52.38171	&1999$-$12$-$30T19:25:10&   51542.8  & 14954.87&       3.69$\times 10^{1}$     \\
653	&300022	&NGC5139	&201.69753	&$-$47.47309	&2000$-$01$-$24T02:15:01&   51567.1  & 25026.24&       9.43    \\
676	&400043	&GROJ0422+32	&65.42895	&32.91309	&2000$-$10$-$10T06:57:08&   51827.3  & 18795.74&       1.66$\times 10^{1}$     \\
677	&400044	&4U1543$-$47	&236.78270	&$-$47.67433	&2000$-$07$-$26T10:30:33&   51751.4  & 9844.94 &       4.00$\times 10^{1}$     \\
799	&600102	&NGC1395	&54.65637	&$-$23.06370	&1999$-$12$-$31T06:31:53&   51543.3  & 18983.21&       1.97    \\
868	&700173	&PG1115+407	&169.67772	&40.42151	&2000$-$10$-$03T23:12:41&   51821.0  & 17338.21&       1.91    \\
887	&700192	&Elais:N2	&249.19620	&41.02638	&2000$-$08$-$02T17:43:11&   51758.7  & 73375.76&       1.07    \\
888	&700193	&Elais:N1	&242.58421	&54.55655	&2000$-$08$-$03T14:45:16&   51759.6  & 71115.65&       1.35    \\
944	&900016	&SGRB2		&266.78070	&$-$28.44160	&2000$-$03$-$29T09:44:36&   51632.4  & 98629.03&       1.26$\times 10^{2}$     \\
945	&900017	&GALACTICCENTERA&266.58221	&$-$28.87193	&2000$-$07$-$07T19:05:19&   51732.8  & 48399.41&       1.10$\times 10^{2}$     \\
949	&900021	&GALACTICPLANE	&280.99377	&$-$4.07476	&2000$-$02$-$25T22:00:10&   51599.9  & 38385.57&       1.88$\times 10^{2}$     \\
953	&300028	&47TUC		&5.97537	&$-$72.07276	&2000$-$03$-$16T08:39:44&   51619.4  & 31676.94&       4.94    \\
955	&300030	&47TUC		&5.97539	&$-$72.07265	&2000$-$03$-$16T18:33:03&   51619.8  & 31676.94&       4.94    \\
957	&900024	&HUBBLEDEEPFIELD&189.15215	&62.24523	&2000$-$02$-$23T06:31:59&   51597.3  & 57442.50&       1.49    \\
966	&900026	&HUBBLEDEEPFIELD&189.26813	&62.21516	&1999$-$11$-$21T04:03:48&   51503.2  & 55972.70&       1.49    \\
967	&900027	&HUBBLEDEEPFIELD&189.26808	&62.21466	&1999$-$11$-$14T19:47:49&   51496.8  & 57433.02&       1.49    \\
972	&200079	&M17		&275.12656	&$-$16.17502	&2002$-$03$-$02T17:05:03&   52335.7  & 39439.75&       1.45$\times 10^{2}$     \\
977	&200084	&NGC6530	&271.10144	&$-$24.35215	&2001$-$06$-$18T11:40:26&   52078.5  & 58008.05&       7.09$\times 10^{1}$     \\
978	&200085	&M16		&274.68353	&$-$13.80290	&2001$-$07$-$30T18:55:00&   52120.8  & 77126.02&       1.24$\times 10^{2}$     \\
1109	&780059	&PKS0312$-$770	&47.99976	&$-$76.86046	&1999$-$09$-$08T02:42:07&   51429.1  & 12836.97&       8.20    \\
1110	&780060	&PKS0312$-$770	&48.13725	&$-$76.84995	&1999$-$09$-$08T06:48:47&   51429.3  & 12581.05&       8.20    \\
1111	&780061	&PKS0312$-$770	&48.35330	&$-$76.83350	&1999$-$09$-$08T10:38:47&   51429.4  & 12593.69&       8.17    \\
1232	&280182	&NGC2516	&119.58367	&$-$60.78894	&1999$-$08$-$27T07:07:36&   51417.3  & 5728.96 &       1.46$\times 10^{1}$     \\
1249	&280199	&ETACARINAE	&161.25478	&$-$59.68141	&1999$-$09$-$06T23:46:37&   51428.0  & 9630.09 &       1.36$\times 10^{2}$     \\
1479	&980429	&LEONIDANTI$-$RADI&333.29935	&$-$22.18443	&1999$-$11$-$17T22:42:28&   51499.9  & 19690.89&       2.45    \\
1519	&300022	&NGC5139	&201.69749	&$-$47.47309	&2000$-$01$-$25T04:33:40&   51568.2  & 43591.34&       9.43    \\
1523	&900021	&GALACTICPLANE	&280.99301	&$-$4.07539	&2000$-$02$-$24T09:56:58&   51598.4  & 54724.64&       1.88$\times 10^{2}$     \\
1671	&900030	&HDFNORTH	&189.26709	&62.21755	&2000$-$11$-$21T13:26:31&   51869.6  & 166146.53&      1.49    \\
1672	&900031	&AXAFSOUTHERNDEE&53.11979	&$-$27.81285	&2000$-$12$-$16T05:07:55&   51894.2  & 95138.09&       9.01$\times 10^{-1}$    \\
1697	&900056	&ISONW\#3	&158.65106	&57.62013	&2001$-$05$-$16T12:46:50&   52045.5  & 38863.00&       5.70$\times 10^{-1}$    \\
1698	&900057	&ISONW\#1	&158.50160	&57.77004	&2001$-$05$-$17T18:29:38&   52046.8  & 72974.50&       5.72$\times 10^{-1}$    \\
1699	&900058	&ISONW\#2	&158.33333	&57.62089	&2001$-$04$-$30T10:59:38&   52029.5  & 40735.17&       5.74$\times 10^{-1}$    \\
1866	&200094	&LYNDS1551	&67.88639	&18.13492	&2001$-$07$-$23T05:10:11&   52113.2  & 78926.84&       1.86$\times 10^{1}$     \\
1867	&200095	&CHAINORTHCLOUD	&167.53415	&$-$76.57798	&2001$-$07$-$02T06:23:33&   52092.3  & 66292.12&       8.62    \\
\noalign{\smallskip}
\hline
\end{tabular}
\end{center}
\end{table*}

\setcounter{table}{0}
\begin{table*}
\begin{center}
\caption{Field Properties in the BMW-C Sample, Continued.}
\begin{tabular}{rllcccccc}
\hline
\hline
\noalign{\smallskip}
ID & Sequence & Target Name & Nominal& Nominal Dec   & Observation Date & MJD  & Effective  & Average  $N_{\rm H}$ \\
  ({\tt OBSID})    & ({\tt SEQ\_NUM})  & ({\tt OBJECT})&  RA ({\tt RA\_NOM}) & ({\tt DEC\_NOM}) & ({\tt DATE\_OBS}) & & Exp.\  & ({\tt NH\_WAVG)} \\ 
               &                 &             & (J2000)       & (J2000)        &      &            & ({\tt TEFF}, s)        & ($10^{20}$ cm$^{-2}$) \\ 
\noalign{\smallskip}
\hline
\noalign{\smallskip}
1873	&200101	&M67		&132.84630	&11.82625	&2001$-$05$-$31T11:19:36&   52060.5  & 46248.46&       3.94    \\
1874	&200102	&ROSETTEFIELD1	&97.97012	&4.92848	&2001$-$01$-$05T11:54:17&   51914.5  & 14450.83&       7.33$\times 10^{1}$     \\
1875	&200103	&ROSETTEFIELD2	&98.17007	&4.71265	&2001$-$01$-$05T17:47:55&   51914.7  & 19506.63&       7.47$\times 10^{1}$     \\
1876	&200104	&ROSETTEFIELD3	&98.32136	&4.57847	&2001$-$01$-$05T23:29:45&   51915.0  & 19408.68&       7.47$\times 10^{1}$     \\
1877	&200105	&ROSETTEFIELD4	&98.57216	&4.46292	&2001$-$01$-$06T05:11:35&   51915.2  & 19509.79&       7.41$\times 10^{1}$     \\
1878	&200106	&NGC\_2024	&85.44305	&$-$1.91993	&2001$-$08$-$08T06:37:10&   52129.3  & 74671.84&       2.14$\times 10^{1}$     \\
1881	&200109	&NGC346		&14.76239	&$-$72.17383	&2001$-$05$-$15T01:55:17&   52044.1  & 98670.55&       6.23    \\
1882	&200110	&MONOCEROSR2	&91.96145	&$-$6.38078	&2000$-$12$-$02T23:14:19&   51881.0  & 95348.23&       2.46$\times 10^{1}$     \\
1893	&200121	&S106FIR	&306.85522	&37.37556	&2001$-$11$-$03T00:08:14&   52216.0  & 44387.49&       1.18$\times 10^{2}$     \\
1934	&400147	&XTEJ1723$-$376	&260.90622	&$-$37.66724	&2001$-$09$-$04T06:40:52&   52156.3  & 28592.80&       1.43$\times 10^{2}$     \\
2232	&900059	&HDF$-$N	&189.14977	&62.24391	&2001$-$02$-$19T14:24:58&   51959.6  & 129242.30&      1.49    \\
2233	&900060	&HDF$-$N	&189.14841	&62.24338	&2001$-$02$-$22T03:44:27&   51962.2  & 62157.43&       1.49    \\
2234	&900061	&HDF$-$N	&189.14441	&62.24145	&2001$-$03$-$02T02:00:29&   51970.1  & 162255.00&      1.49    \\
2235	&900062	&XMM13HR$-$FIELD4&203.57935	&37.82424	&2001$-$06$-$08T10:37:06&   52068.4  & 30002.42&       8.49$\times 10^{-1}$    \\
2236	&900063	&XMM13HR$-$FIELD1&203.55498	&37.97983	&2001$-$06$-$08T19:25:34&   52068.8  & 29800.20&       8.31$\times 10^{-1}$    \\
2237	&900064	&XMM13HR$-$FIELD2&203.77635	&37.84336	&2001$-$06$-$09T04:01:09&   52069.2  & 29797.05&       8.47$\times 10^{-1}$    \\
2238	&900065	&XMM13HR$-$FIELD3&203.75237	&37.99891	&2001$-$06$-$09T12:36:44&   52069.5  & 18884.53&       8.32$\times 10^{-1}$    \\
2239	&900066	&CDFS		&53.11699	&$-$27.81122	&2000$-$12$-$23T17:28:01&   51901.7  & 129740.39&      9.00$\times 10^{-1}$    \\
2240	&900067	&CADIS01HFIELD	&26.90029	&2.32900	&2001$-$01$-$26T18:36:47&   51935.8  & 26702.29&       3.03    \\
2252	&900079	&WHDF		&5.63854	&0.34335	&2001$-$01$-$06T11:36:05&   51915.5  & 71220.99&       2.67    \\
2254	&900081	&3C295		&212.82706	&52.20298	&2001$-$05$-$18T15:25:59&   52047.6  & 89236.41&       1.33    \\
2255	&900082	&NGC55		&3.79150	&$-$39.22025	&2001$-$09$-$11T06:25:05&   52163.3  & 58576.13&       1.72    \\
2267	&900094	&GCS20		&266.17123	&$-$29.27356	&2001$-$07$-$19T10:01:48&   52109.4  & 10208.63&       1.19$\times 10^{2}$     \\
2268	&900095	&GCS21		&265.98087	&$-$29.17148	&2001$-$07$-$20T04:37:11&   52110.2  & 10811.74&       1.16$\times 10^{2}$     \\
2270	&900097	&GCS22		&266.24515	&$-$29.54168	&2001$-$07$-$20T08:00:49&   52110.3  & 10622.17&       1.23$\times 10^{2}$     \\
2272	&900099	&GCS23		&266.05414	&$-$29.43966	&2001$-$07$-$20T11:12:40&   52110.5  & 11611.05&       1.26$\times 10^{2}$     \\
2273	&900100	&GCS10		&266.71024	&$-$28.87570	&2001$-$07$-$18T00:48:28&   52108.0  & 11611.05&       1.11$\times 10^{2}$     \\
2274	&900101	&GCS3		&266.67642	&$-$28.17316	&2001$-$07$-$16T08:44:25&   52106.4  & 10426.28&       1.29$\times 10^{2}$     \\
2275	&900102	&GCS24		&265.86353	&$-$29.33749	&2001$-$07$-$20T14:41:10&   52110.6  & 11598.25&       1.22$\times 10^{2}$     \\
2276	&900103	&GCS11		&266.52002	&$-$28.77435	&2001$-$07$-$18T04:16:58&   52108.2  & 11520.58&       1.13$\times 10^{2}$     \\
2277	&900104	&GCS4		&266.94061	&$-$28.54206	&2001$-$07$-$16T11:52:55&   52106.5  & 10426.28&       1.22$\times 10^{2}$     \\
2278	&900105	&GCS25		&266.12796	&$-$29.70796	&2001$-$07$-$20T18:09:40&   52110.8  & 10030.49&       1.30$\times 10^{2}$     \\
2279	&900106	&GCS12		&266.33029	&$-$28.67280	&2001$-$07$-$18T07:45:28&   52108.3  & 11611.05&       1.10$\times 10^{2}$     \\
2280	&900107	&GCS5		&266.75079	&$-$28.44106	&2001$-$07$-$16T15:01:25&   52106.6  & 10426.28&       1.26$\times 10^{2}$     \\
2281	&900108	&GCS26		&265.93677	&$-$29.60579	&2001$-$07$-$20T21:38:10&   52110.9  & 8219.31 &       1.32$\times 10^{2}$     \\
2282	&900109	&GCS13		&266.59457	&$-$29.04224	&2001$-$07$-$18T11:13:58&   52108.5  & 10625.33&       1.07$\times 10^{2}$     \\
2283	&900110	&GCS27		&265.74591	&$-$29.50336	&2001$-$07$-$21T01:06:39&   52111.0  & 11614.25&       1.29$\times 10^{2}$     \\
2284	&900111	&GCS14		&266.40414	&$-$28.94089	&2001$-$07$-$18T14:25:48&   52108.6  & 10625.33&       1.09$\times 10^{2}$     \\
2285	&900112	&GCS6		&266.56137	&$-$28.33985	&2001$-$07$-$16T18:09:55&   52106.8  & 10410.49&       1.22$\times 10^{2}$     \\
2286	&900113	&GCS28		&266.01025	&$-$29.87408	&2001$-$07$-$21T04:35:09&   52111.2  & 11614.25&       1.37$\times 10^{2}$     \\
2287	&900114	&GCS15		&266.21411	&$-$28.83905	&2001$-$07$-$18T17:37:38&   52108.7  & 10625.33&       1.07$\times 10^{2}$     \\
2288	&900115	&GCS7		&266.82550	&$-$28.70883	&2001$-$07$-$17T14:11:51&   52107.6  & 11493.24&       1.17$\times 10^{2}$     \\
2289	&900116	&GCS29		&265.81894	&$-$29.77175	&2001$-$07$-$21T08:03:39&   52111.3  & 11614.25&       1.40$\times 10^{2}$     \\
2290	&900117	&GCS30		&265.62793	&$-$29.66920	&2001$-$07$-$21T11:32:10&   52111.5  & 11614.25&       1.36$\times 10^{2}$     \\
2291	&900118	&GCS16		&266.47845	&$-$29.20889	&2001$-$07$-$18T20:49:28&   52108.9  & 10625.33&       1.11$\times 10^{2}$     \\
2292	&900119	&GCS8		&266.63556	&$-$28.60780	&2001$-$07$-$17T17:51:28&   52107.7  & 11604.73&       1.19$\times 10^{2}$     \\
2293	&900120	&GCS17		&266.28790	&$-$29.10730	&2001$-$07$-$19T00:01:18&   52109.0  & 11118.21&       1.13$\times 10^{2}$     \\
2294	&900121	&GCS9		&266.44598	&$-$28.50635	&2001$-$07$-$17T21:19:58&   52107.9  & 11611.05&       1.16$\times 10^{2}$     \\
2295	&900122	&GCS18		&266.09775	&$-$29.00538	&2001$-$07$-$19T03:21:28&   52109.1  & 11118.21&       1.10$\times 10^{2}$     \\
2296	&900123	&GCS19		&266.36200	&$-$29.37541	&2001$-$07$-$19T06:41:38&   52109.3  & 11118.21&       1.17$\times 10^{2}$     \\
2298	&900125	&GALACTICPLANE1	&280.88382	&$-$3.90722	&2001$-$05$-$20T08:44:32&   52049.4  & 94214.80&       2.00$\times 10^{2}$     \\
2300	&900127	&1RXSJ104047.4$-$7&160.18372	&$-$70.78441	&2001$-$09$-$11T02:20:34&   52163.1  & 12802.17&       1.05$\times 10^{1}$     \\
2312	&900066	&CDFS		&53.11796	&$-$27.81074	&2000$-$12$-$13T03:28:03&   51891.1  & 123686.25&      9.00$\times 10^{-1}$    \\
2313	&900066	&CDFS		&53.11699	&$-$27.81122	&2000$-$12$-$21T02:08:43&   51899.1  & 130389.61&      9.00$\times 10^{-1}$    \\
\noalign{\smallskip}
\hline
\end{tabular}
\end{center}
\end{table*}

\setcounter{table}{0}
\begin{table*}
\begin{center}
\caption{Field Properties in the BMW-C Sample, Continued.}
\begin{tabular}{rllcccccc}
\hline
\hline
\noalign{\smallskip}
ID & Sequence & Target Name & Nominal& Nominal Dec   & Observation Date & MJD  & Effective  & Average  $N_{\rm H}$ \\
  ({\tt OBSID})    & ({\tt SEQ\_NUM})  & ({\tt OBJECT})&  RA ({\tt RA\_NOM}) & ({\tt DEC\_NOM}) & ({\tt DATE\_OBS}) & & Exp.\  & ({\tt NH\_WAVG)} \\ 
               &                 &             & (J2000)       & (J2000)        &      &            & ({\tt TEFF}, s)        & ($10^{20}$ cm$^{-2}$) \\ 
\noalign{\smallskip}
\hline
\noalign{\smallskip}
2344	&900030	&HDFNORTH	&189.26715	&62.21756	&2000$-$11$-$24T05:41:51&   51872.2  & 66696.32&       1.49    \\
2386	&900030	&HDFNORTH	&189.26720	&62.21756	&2000$-$11$-$20T05:39:48&   51868.2  & 9842.40 &       1.49    \\
2405	&900031	&AXAFSOUTHERNDEE&53.12021	&$-$27.81259	&2000$-$12$-$11T08:14:17&   51889.3  & 56702.80&       9.01$\times 10^{-1}$    \\
2406	&900066	&CDFS		&53.11819	&$-$27.81062	&2000$-$12$-$10T23:35:12&   51889.0  & 29686.29&       9.00$\times 10^{-1}$    \\
2409	&900066	&CDFS		&53.11699	&$-$27.81123	&2000$-$12$-$19T03:55:35&   51897.2  & 68982.46&       9.00$\times 10^{-1}$    \\
2421	&900061	&HDF$-$N	&189.14301	&62.24069	&2001$-$03$-$04T16:52:28&   51972.7  & 61648.85&       1.49    \\
2423	&900060	&HDF$-$N	&189.14841	&62.24339	&2001$-$02$-$23T06:57:57&   51963.3  & 68419.52&       1.49    \\
2550	&200158	&NGC2264	&100.19905	&9.84542	&2002$-$02$-$09T05:10:47&   52314.2  & 48137.83&       4.70$\times 10^{1}$     \\
2553	&200161	&MADDALENA'SCLOU&102.27326	&$-$4.56858	&2002$-$02$-$08T20:39:21&   52313.9  & 20757.27&       7.54$\times 10^{1}$     \\
2556	&200164	&RCW38		&134.83614	&$-$47.50389	&2001$-$12$-$10T10:14:38&   52253.4  & 95599.90&       8.93$\times 10^{1}$     \\
2562	&200170	&IRAM04191	&65.48610	&15.49148	&2002$-$02$-$09T19:08:44&   52314.8  & 18918.96&       1.80$\times 10^{1}$     \\
3293	&900132	&HDF$-$N	&189.21547	&62.21753	&2001$-$11$-$13T15:04:42&   52226.6  & 161250.19&      1.49    \\
3294	&900133	&HDF$-$N	&189.15369	&62.24493	&2002$-$02$-$14T03:22:07&   52319.1  & 166971.55&      1.49    \\
3305	&900144	&GROTH$-$WESTPHALF&214.42932	&52.47367	&2002$-$08$-$11T21:43:57&   52497.9  & 29405.28&       1.28    \\
3343	&900182	&LH$-$NW$-$6	&158.33452	&57.92092	&2002$-$05$-$03T09:11:41&   52397.4  & 31812.35&       5.91$\times 10^{-1}$    \\
3344	&900183	&LH$-$NW$-$7	&158.18425	&57.77091	&2002$-$05$-$01T20:03:06&   52395.8  & 38545.61&       5.89$\times 10^{-1}$    \\
3345	&900184	&LH$-$NW$-$4	&158.01767	&57.62109	&2002$-$04$-$29T03:23:45&   52393.1  & 38466.62&       5.89$\times 10^{-1}$    \\
3346	&900185	&LH$-$NW$-$5	&158.50150	&57.47105	&2002$-$04$-$30T02:03:59&   52394.1  & 38207.51&       5.73$\times 10^{-1}$    \\
3347	&900186	&LH$-$NW$-$8	&158.65106	&57.92095	&2002$-$05$-$02T14:16:27&   52396.6  & 38463.43&       5.72$\times 10^{-1}$    \\
3348	&900187	&LH$-$NW$-$9	&158.80956	&57.77088	&2002$-$05$-$04T11:01:47&   52398.5  & 39521.85&       5.69$\times 10^{-1}$    \\
3388	&900132	&HDF$-$N	&189.21545	&62.21752	&2001$-$11$-$16T05:41:16&   52229.2  & 49560.66&       1.49    \\
3389	&900132	&HDF$-$N	&189.21523	&62.21782	&2001$-$11$-$21T14:16:05&   52234.6  & 102258.35&      1.49    \\
3390	&900133	&HDF$-$N	&189.15369	&62.24493	&2002$-$02$-$16T18:58:56&   52321.8  & 161733.11&      1.49    \\
3391	&900133	&HDF$-$N	&189.15369	&62.24493	&2002$-$02$-$22T01:52:24&   52327.1  & 164732.22&      1.49    \\
3408	&900132	&HDF$-$N	&189.21547	&62.21753	&2001$-$11$-$17T01:09:01&   52230.0  & 66213.38&       1.49    \\
3409	&900133	&HDF$-$N	&189.15367	&62.24494	&2002$-$02$-$12T10:59:50&   52317.5  & 82233.02&       1.49    \\
4357	&900144	&GROTH$-$WESTPHALF&214.42932	&52.47367	&2002$-$08$-$12T22:32:00&   52498.9  & 84361.27&       1.28    \\
4365	&900144	&GROTH$-$WESTPHALF&214.42934	&52.47367	&2002$-$08$-$21T10:56:53&   52507.5  & 58854.59&       1.28    \\
\noalign{\smallskip}
\hline
\end{tabular}
\end{center}
\end{table*}

	\section{Data processing\label{bmwc:dataproc}} 

Our pipeline is a combination of {\tt Ciao}\footnote{http://cxc.harvard.edu/ciao/} 
tasks (data screening,  
image reduction, exposure map creation), IDL programs (additional data 
screening, wavelet detection core), tasks in the 
{\tt HEAsoft} 
package and UNIX shell scripts (drivers and house-keeping) that reduces and 
analyses the Level 2 (L2) data generated by the {\it Chandra} X-ray Center 
(CXC)  standard data processing in a uniform fashion.
The data (event list, the aspect solution, the aspect offset, and the 
bad pixel files) were downloaded using the {\it Chandra} Search and Retrieval 
tool (ChaSeR) from the {\it Chandra} Data Archive 
(CDA)\footnote{http://cxc.harvard.edu/cda/s+r.html.}, and 
were filtered to only include the standard event grades 
[Advanced Satellite for Cosmology and Astrophysics ({\it ASCA}) grades 0, 2, 3, 4, 6]. 
Since several {\it Chandra} data sets in the archive suffer from known 
aspect offsets as large as 2$\arcsec$, we checked and corrected 
for this problem using the {\tt fix\_batch} Perl script by 
Tom Aldcroft\footnote{http://cxc.harvard.edu/cal/ASPECT/fix\_offset/fix\_offset.cgi.}.

We applied energy filters to the offset-corrected L2 event list, 
and created soft (SB, 0.5--2.0\,keV), hard (HB, 2.0--7.0\,keV) and total 
(FB, 0.5--7.0\,keV) band event files. 
The upper limit on our hard and total energy bands was chosen at 
7\,keV because at higher energy the background increases and the effective 
area decreases, producing lower  signal-to-noise (S/N) data.
Our results in the 0.5--10\,keV band are then extrapolations 
from our findings in the 0.5--7\,keV range 
(see Sects.~\ref{bmwc:catalogue}, and \ref{bmwc:fluxes}).
	
\begin{figure}	
	 	\resizebox{\hsize}{!}{\includegraphics{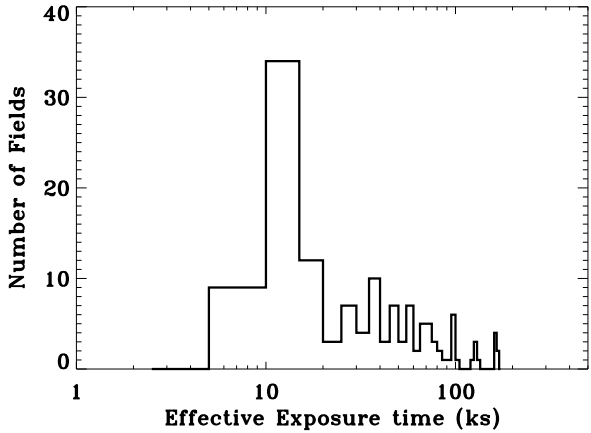}}
 		\caption{Distribution of effective exposure 
		times for the 136 observations.}
 		\label{bmwc:histo_teff}
	\end{figure}

Since the source detection strongly depends on the background rate, 
the data obtained during background flares were carefully removed.  
We created counts light curves with a time resolution that would allow  
$\sim 400$ counts per time bin, which typically correspond to  
$\sim 1.4$\,ks and $\sim 1.8$\,ks in the soft and hard bands, respectively,  
and excluded all time intervals during which the rate was more than  
3$\sigma$ of the mean rate. 
The mean fraction of exposure lost to background flares is  4\%, 
comparable with 5\% given by the ChaMP 
collaboration for FI chips (\citealt{Champ3,Champ1}). 
The effective exposure times reported in the catalogue 
({\tt TEFF}, see Table~\ref{bmwc:fieldprops} and distribution 
shown in Fig.~\ref{bmwc:histo_teff}, median 31.8\,ks), 
reflect these corrections.

Generally, hot/flickering pixels and bad columns are listed in the 
calibration database and eliminated by the {\tt Ciao} tools. 
However, some remaining flickering pixels, probably due to the afterglow of 
cosmic rays or charged particles hitting the CCDs, occur and they 
were identified at this point. 
Following \citet{Tozziea01}, we defined as flickering each pixel that 
registered two events within the time of two consecutive frames. 
In each observation we eliminated all the events registered from that pixel,
a procedure which is safe when no bright sources are present, such as is 
our case. 

	\subsection{Total background map creation\label{bmwc:bgdmaps}} 
	%
The wavelet algorithm requires an accurate characterisation of the background
of the images to process (see Sec.~\ref{bmwc:algorithm}), 
since the WT detection is carried 
out on top of a map representing the image background 
(the total background map; \citealt{Lazzatiea99}). 
The background in the ACIS-I images, in particular, is the sum of two components, 
a cosmic X-ray background
and a particle one (i.e., cosmic--ray-induced).
The former suffers from vignetting and can be modelled 
as a power law of photon index $\Gamma =1.4$ 
(\citealt{Marshallea80}; 
\citealt{Morettiea03}).
The latter is not affected by vignetting, only depends on the temperature 
of the electronics, and can be modelled as a flat pattern with an inaccuracy within 
10\%\footnote{http://cxc.harvard.edu/cal/Links/Acis/acis/Cal\_prods/bkgrnd/current/.}.
For each field and for each of the three energy bands, 
we used the {\tt merge\_all} script to create 
exposure maps (adopting an input spectrum of a power law with photon index 1.4) 
and flat maps and then combined them.

	\section{Source detection and catalogue construction\label{bmwc:detect}} 

	\subsection{The BMW-C algorithm \label{bmwc:algorithm}} 

We used an updated version of the 
BMW detection algorithm  that supports the analysis of {\it Chandra} ACIS images. 
The main steps of the algorithm can be summarised as follows 
(full details in \citealt{Lazzatiea98}; \citealt{Lazzatiea99};  \citealt{Campanaea99}).
The first step is the creation of the WT of the input image; the BMW WT 
is based on the discrete multi-resolution theory and on the ``\`a{} trous'' algorithm
(\citealt{trous}), which differs from continuous--WT-based algorithms which can 
sample more scales at the cost of a longer computing time 
(\citealt{Rosatiea95,Grebenevea95,Damianiea97a}). 
We used a Mexican hat mother-wavelet, which can be analytically approximated 
by the difference of two Gaussian functions (\citealt{Slezakea94}). 
The WT decomposes each image into a set of sub-images, each of them carrying
the information of the original image at a given scale. This property  
makes the WT well suited for the analysis of X-ray images,
where the scale of sources is not constant over the field of view,
because of the dependence of the PSF on the off-axis angle.
We used 7 WT scales $a=[1,2,4,8,16,32,64]$ pixels 
to cover a wide range of source sizes, where $a$ is the scale of the WT
(\citealt{Lazzatiea99}).

Candidate sources are identified as local maxima above the significance threshold 
in the wavelet space at each scale, so that a list is obtained at each scale, 
and then a cross-match is performed among the 7 lists to merge them.
At the end of this step, we have a first estimate of source positions
(the pixel with the highest WT coefficient),  
source counts (the local maximum of the WT) 
and a guess of the source extension (the scale at which the WT is maximized). 
A critical parameter is the detection threshold which, in the context of WT algorithms,
is usually fixed arbitrarily by the user in terms of expected spurious detections 
per field (\citealt{Lazzatiea98}). 
The number of expected spurious detections as a function of the threshold 
value and for each scale was calculated 
by means of Monte Carlo simulations (\citealt{Morettiea02}). 
We ran the detection algorithm with a single significance threshold 
that corresponds to $\sim 0.1$ spurious detections per scale, hence (with 
7 scales) $\sim 0.7$ spurious detections per field for each band in which we performed the detection. 
Given our sample of 136 fields, we expect a total of $\sim 95$ spurious 
sources in the catalogue. 

The final step is the characterisation of the sources by means of a multi-source 
$\chi ^2$ minimization with respect to a Gaussian model source in the WT space. 
In order to fit the model on a set of independent data, the WT coefficients are 
decimated according to a scheme described in full in Lazzati et al.\ (1999). 

\begin{figure}
         \resizebox{\hsize}{!}{\includegraphics{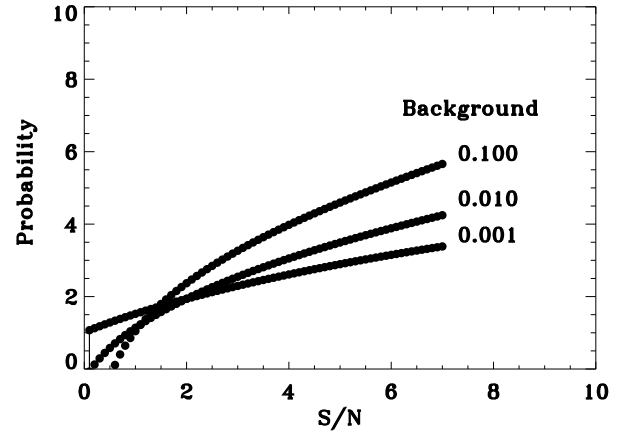}}
         \caption{ Wavelet probability (or source significance in number of $\sigma$s) 
          as a function of WT S/N and background value (counts, see Sect.~\ref{bmwc:algorithm}).     }
         \label{bmwc:cha_prob}
        \end{figure}

The wavelet probability  is defined as the
confidence level at which a source is not a chance background 
fluctuation, given the background statistics and the specific field 
exposure time.  
This quantity is assessed via the S/N ratio in WT space. 
For each WT scale, the noise level is computed through numerical
simulations of blank fields with the corresponding background 
(\citealt{Lazzatiea99}), while the signal is the peak of the WT 
coefficients corresponding to the source. 
Figure~\ref{bmwc:cha_prob} shows the wavelet probability as a 
function of WT S/N, and background value.
In order to make this significance more easily
comparable to other methods, confidence intervals are approximately expressed in
units of the standard deviation $\sigma$ for which a Gaussian
distributed variable would give an equal probability of spurious
detection (68\%: $1\,\sigma$; 95\%: $2\,\sigma$, etc.). 
So, the values of the wavelet probability 
in the catalogue column {\tt WTPROBAB} represent the number of 
$\sigma$'s corresponding to the confidence level of that specific source.  

       \setcounter{figure}{3}
	\begin{figure*}	
	\centering
	 	\includegraphics[width=19cm,height=8.5cm]{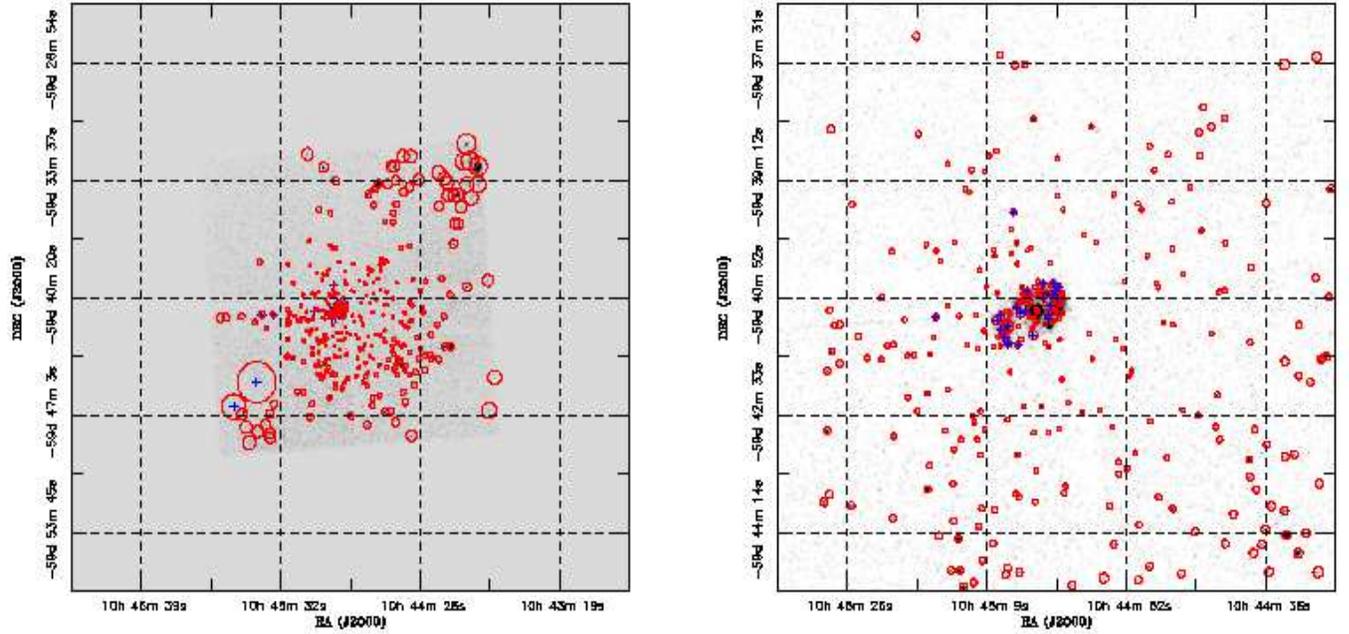}
 		\caption[Detection Results]{Example of Detection. {\bf Left}: the $\eta$ Carinae 
		full field ({\tt OBSID} 50, {\tt SEQ\_NUM} 200019) at half resolution. 
		The FOV is $\sim 33\arcmin$. The sources are represented 
		with a size which is three times their Gaussian width.
		Note the complicated extended structure at the centre and 
		the spurious detections along a readout streak (see Sect.~\ref{bmwc:algorithm})
		Crosses mark sources that the detection algorithm classifies as extended 
		(e.g.\ bottom-left corner).
		 {\bf Right}: central portion of the field at full resolution. 
		The FOV is $\sim 8\farcm4$ and the sources are represented with a size 
			which is five times their Gaussian width. }
 		\label{bmwc:origimas}
	\end{figure*}
	\begin{figure*}	
	\centering
	 	\includegraphics[width=19cm,height=8.5cm]{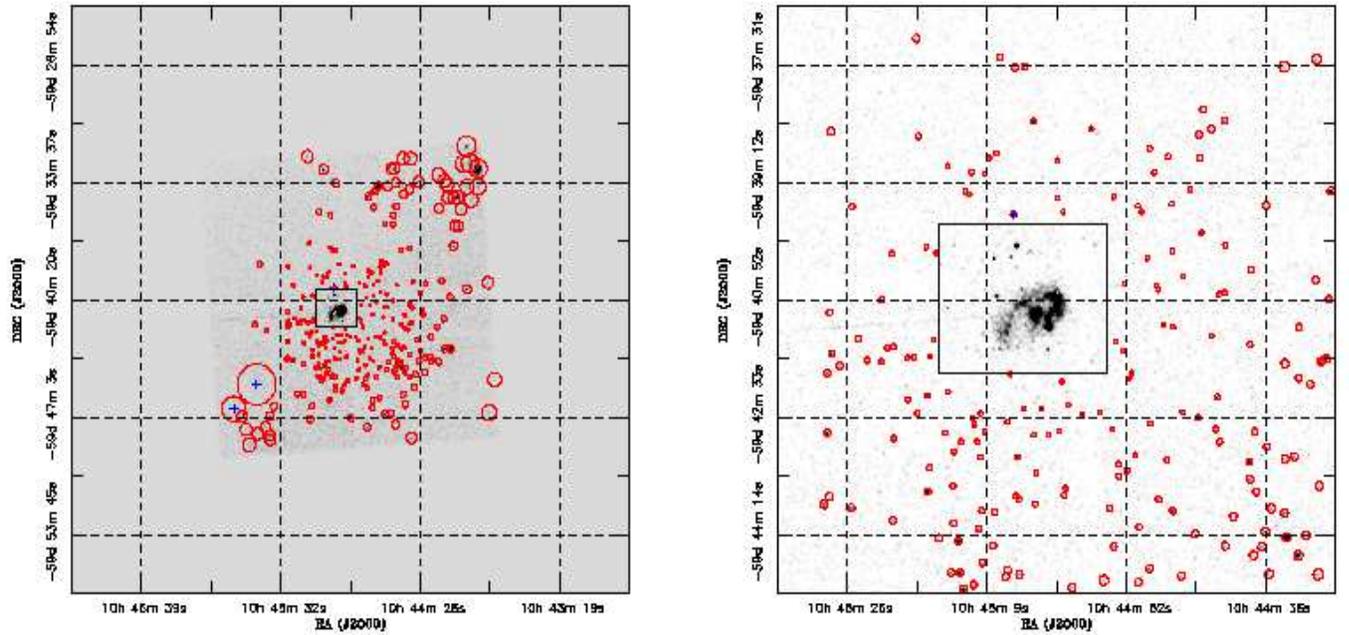}
 		\caption[Detection Results 2]{Example of manual cleaning. 
		The spurious sources along the readout streak were eliminated and the 
		sources in the central portion of the image 
		(contained within the box and not shown) were flagged.}
 		\label{bmwc:cleanimas}
	\end{figure*}

We ran the detection algorithm on the central $\sim 8\arcmin$ portion of the 
source image at full resolution (rebin $=1$, 1 pixel $\sim 0\farcs49$)  
and on the image rebinned by a factor of two outside (rebin $=2$, 1 pixel $\sim 0\farcs98$). 
This strategy allowed us to optimize computer time, while preserving spatial  
information. Indeed, outside the $\sim 8\arcmin$ radius circle,   
the PSF radius that encircles 90\% of the energy is $\ga 6\farcs5$, i.e., $\ga6.6$  
rebinned pixels in the soft band  
and $\ga 7\farcs4$, i.e., $\ga7.5$ rebinned pixels in the hard band. 
In detail,   
a) we ran the detection on the images at rebin 2;   
b) we ran the detection in their inner $512\times512$ part  
	(rebinned pixels, corresponding to $\sim 8\farcm4 \times 8\farcm4$) 
	at the full resolution;  
c) we excluded the $480\times480$ pixel central part in the analysis at rebin 2  
	(rebinned pixels, corresponding to $\sim 7\farcm8\times 7\farcm8$); 
d) we cross-matched the positions of the sources found at rebin 1 and 2 to  
	exclude common double entries (sources were considered coincident if their  
	distance was less than 6 times their width).
We repeated this procedure for each of the three energy bands, and  
cross-matched the resulting source coordinates to form the definitive list
(for coincident sources, the coordinates of the highest S/N one were kept).

A source extension, defined as the width of the best-fitting Gaussian model,  
was calculated for each band and the extension  
of the highest S/N among the three bands and the inner and outer regions was kept. 

An example of the products of the pipeline is shown in  
Figs.~\ref{bmwc:origimas} and \ref{bmwc:cleanimas}, which represent the 
$\eta$ Carinae field ({\tt OBSID} 50, {\tt SEQ\_NUM} 200019), one of the most problematic
observations since it has a complicated extended structure at the centre as well as 
other data problems. 
Figure~\ref{bmwc:origimas} (left) is the full image at rebin 2, with a 
FOV of $\sim 33\arcmin$.  The sources are represented 
with circles with a size which is three times their Gaussian width. 
Note the spurious detections along a readout streak (photons detected during 
the readout in a field containing a bright source are clocked out in the wrong row 
and so have incorrect CHIPY values and show up as a streak along the column). 
Crosses mark sources that the detection algorithm classifies as extended.
Figure~\ref{bmwc:origimas} (right) shows the central portion of the field at full 
resolution. The FOV is $\sim 8\farcm4$, and the sources are represented with a size 
which is five times their Gaussian width for better presentation.

	\subsection{BMW-C catalogue\label{bmwc:catalogue}} 
The pipeline produced a catalogue of source positions, count rates, 
counts, extensions, and relative errors, as well as the additional 
information drawn from the headers of the original files for a total of 21325 sources.

Source counts were corrected for vignetting using the 
exposure maps (see Sect.~\ref{bmwc:bgdmaps}),  
i.e., by comparing the image counts at each source position with the average counts 
of the normalized exposure map within a PSF range. Furthermore, the source counts were 
corrected for PSF fitting, i.e., for using a Gaussian approximation of the PSF to fit 
the sources in the WT space. This latter correction was calculated by running the detection 
on the psfsize table in {\tt CALDB} (psfsize\_2000830.fits), 
using the PSF images at $\sim 1.50$\,keV for the soft band PSF 
and at $\sim 4.51$\,keV for the hard band PSF, and calculating the percentage of counts lost 
with the Gaussian approximation in the WT detection.

Errors were calculated in two ways: following \citet{Grebenevea95}, appropriate 
for sufficiently high values of the background and source counts 
($\ga 5\times 10^{-2}$ counts/pixel), and 
using basic statistic expressions. 
In the latter case, assuming a Gaussian shaped distribution of $N$ photons 
of width ({\tt WIDTH}) $\Delta$,
the positional errors in $x$ and $y$ ({\tt X\_POS\_E, Y\_POS\_E}) are limited by 
$\sigma_x \approx \sigma_y \approx \Delta / \sqrt{N -1}$, 
the total positional error ({\tt T\_POS\_ER}) is 
$\sigma_t \approx \sqrt{{{\sigma_x}^2} + {{\sigma_y}^2}}$, 
the error on the total number of counts is given by Poisson statistics 
$\sigma_N \approx \sqrt{N}$, 
and the intrinsic limit on the width estimate is 
$\sigma_{\Delta} \approx 0.5 \, \Delta / \sqrt{N-1}$
(see detailed discussion in \citealt{Lazzatiea99}). 
The largest of the values derived using the two methods is adopted. 
As an example, the distribution of the total positional error is shown 
in Fig.~\ref{bmwc:poserr} for the 16834 sources 
that do not require a more in-depth, non-automated analysis (see below). 
%
%
We note that the absolute source position in {\it Chandra} observations 
has an uncertainty radius of 0\farcs6 at 90\% confidence level (CL), and of 0\farcs8 at 
99\% CL\footnote{The most updated information on the {\it Chandra} absolute positional 
accuracy can be found at http://cxc.harvard.edu/cal/ASPECT/celmon/.}. 
These values were calculated by measuring the distances between the {\it Chandra} X-ray source 
positions and the corresponding optical/radio counterpart positions from the Tycho2 
\citep{Tycho2000} or the International Celestial Reference Frame \citep[ICRS, ][]{ICRS}
catalogues, for sources within 3\,$\arcmin$ from the aimpoint 
and with the Science Instrument Module Z-axis (SIM-Z) 
at the nominal detector value (for large off-nominal SIM-Z, the 
observations can suffer an additional aspect offset of up to 0\farcs5).

For each source we computed two fluxes in the 0.5--10\,keV range, assuming that 
a) the source is extragalactic and its flux was corrected for Galactic  
	absorption ({\tt FLUX1}) and  
b) the source is subject to no absorption ({\tt FLUX2}). 
In general, the count-rate--to--flux conversion factors (CF) are a function of 
the assumed shape of the spectrum of the source.
To obtain {\tt FLUX1}, we first calculated an average line-of-sight column density 
$N_{\rm H}$ according to \citet{DickeyL90} with the {\tt nH} tool in {\tt HEASoft} 
({\tt NH\_WAVG}), for each catalogue field. 
We show the distribution of column densities in Fig.~\ref{bmwc:histo_nh}. 
Then we performed simulations with {\tt XSPEC} (\citealt{XSPEC}) 
assuming as input model an absorbed ({\tt wabs}) power law with 
photon index 2.0 (a Crab spectrum, for direct comparison with other 
work, e.g., the BMW-HRI catalogue; \citealt{Panzeraea03}) 
for a range of column densities spanning these values. 
We report the model 0.5--7.0\,keV count rate to 0.5--10\,keV flux 
CFs in the catalogue column {\tt COR\_FAC1}, 
while the flux was calculated as 
{\tt FLUX1} = {\tt COR\_FAC1} $\times$ {\tt T\_FT\_CTS} $/$ {\tt EXPOSURE},
where {\tt T\_FT\_CTS} are the WT total counts and {\tt EXPOSURE} is the total 
exposure time. 
To calculate {\tt FLUX2} an unabsorbed ($N_{\rm H}=0$) spectrum was simulated, and 
the 0.5--7.0\,keV count rate to 0.5--10\,keV flux conversion factor 
is CF2$=8.23\times10^{-12}$ erg cm$^{-2}$ cts$^{-1}$.

For comparison with other work, we also report approximate fluxes in the three bands 
({\tt T\_FLUX}, {\tt S\_FLUX}, and {\tt H\_FLUX},) calculated using a count rate to flux 
conversion factor CF0 $=1\times10^{-11}$ erg cm$^{-2}$ cts$^{-1}$.

       \setcounter{figure}{5}
	\begin{figure} 	
	 \resizebox{\hsize}{!}{\includegraphics{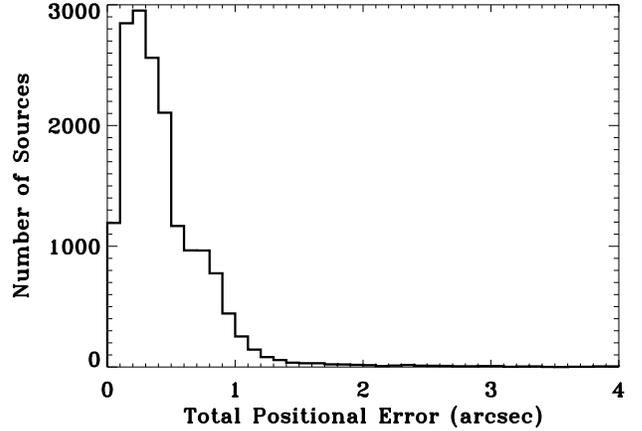}}
	 \caption{Distribution of the total positional error for the 
	16834 BMW-{\it Chandra} sources that do not require a 
	non-automated analysis. The mean is $0\farcs45$ and 
	the median $0\farcs36$.}
	 \label{bmwc:poserr}
	\end{figure}
	\begin{figure}
	 \resizebox{\hsize}{!}{\includegraphics{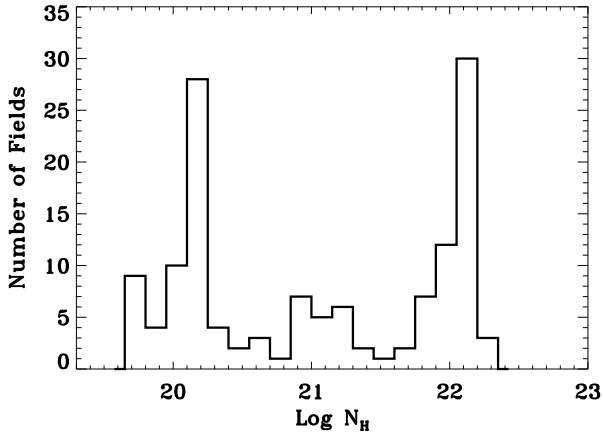}}
	 \caption{Distribution of the Galactic column density of the 
		136 observations according to \citet{DickeyL90}.}
	 \label{bmwc:histo_nh}
	\end{figure}
\begin{table}
\begin{center}
\caption{Energy Bands and Background values.}
\label{bmwc:band_bgd}
\begin{tabular}{lcc}
\hline
\hline
\noalign{\smallskip}
Band & Energy & Background value$^{\mathrm{a}}$ \\
    & (keV) & (counts s$^{-1}$ chip$^{-1}$) \\
\noalign{\smallskip}
\hline
\noalign{\smallskip}
SB  & 0.5-2.0 & ... \\
HB  & 2.0-7.0 & ... \\
FB  & 0.5-0.7 & ... \\
SB1 & 0.5-1.0 & 0.0336  \\
SB2 & 1.0-2.0 & 0.0417  \\
HB1 & 2.0-4.0 & 0.0508  \\
HB2 & 4.0-7.0 & 0.0656  \\
\noalign{\smallskip}
\hline
\end{tabular}
\end{center}
\begin{list}{}{}
\item[$^{\mathrm{a}}$] ACIS Background Memos: 
	Baganoff (1999); Markevitch (2001).
\end{list}
\end{table}

%
%
Each field was visually inspected to identify and remove 
obviously spurious detections (such as image streaks in correspondence 
of bright pointed sources and crowded fields, an example of which is shown 
in Figs.~\ref{bmwc:origimas} and \ref{bmwc:cleanimas}). 
A number of checks were performed on the raw catalogue and the results 
stored in the {\tt FL\_CHECK} catalogue column, as follows.
	\begin{enumerate}
	\item We eliminated sources that presented negative fitted counts
		(an instance that may occur if the WT fit does not converge).
	\item When the errors on position, the total positional error, 
		and the error on the width 
		({\tt X\_POS\_ER, Y\_POS\_ER, T\_POS\_ER, WIDTH\_ER}, respectively),  
		did not converge, they were set to their statistical value 
		({\tt FL\_CHECK $= -2$}), 
		if possible, otherwise the sources were eliminated.
	\item When the error on the width exceeded the width value ({\tt WIDTH}),
		it was set to the statistical value 
		({\tt FL\_CHECK $= -2$}), 
		if possible, otherwise it was left unmodified 
		({\tt FL\_CHECK $= -3$}). 
	\item When the error on the rate ({\tt T\_CR\_ER, S\_CR\_ER, H\_CR\_ER})
        	exceeded the rate value ({\tt T\_CTRATE, S\_CTRATE, H\_CTRATE}),
		it was set to its statistical value 
		({\tt FL\_CHECK $= -2$}), 
		if possible, otherwise it was left unmodified ({\tt FL\_CHECK $= -4$}). 
	\item When the fitted counts ({\tt T\_FT\_CTS, S\_FT\_CTS, H\_FT\_CTS}), 
		and count rates ({\tt T\_CTRATE, S\_CTRATE, H\_CTRATE}),
		and consequently, the fluxes ({\tt FLUX1, FLUX2}), 
		resulted infinite (because the WT fit did not converge), 
		they were set to $-9999.9$.
	\end{enumerate}
Each field was visually inspected again to exclude/flag 
problematic portions, and the results 
stored in the {\tt FL\_CHCK2} catalogue column. In particular, extended pointed 
targets were flagged ({\tt FL\_CHCK2 $= -1$}). 
Sources within a radius of 30$\arcsec$ from the target position 
(as given by {\tt RA\_TARG} and {\tt DEC\_TARG} original header fields) 
of all fields not in surveys (73) were flagged with {\tt FL\_CHCK2 $= 10 \times $FL\_CHCK2}.

The parameters of the final catalogue are listed in Table~\ref{bmwc:params}.
The full catalogue contains 21325 sources, 
16834 of which do not require a more in-depth, non-automated analysis 
(i.e.\ not associated with bright and/or extended sources at the centre of the field).
The latter are obtained applying the following conditions,
{\tt FL\_CHCK2 $\neq -1$}, and  {\tt FL\_CHCK2 $\neq -10$}, and include the pointed ones. 
The final number of sources not associated with pointed targets is 16758 
(see Sect.~\ref{bmwc:ssc}).

\setcounter{table}{2}
\begin{table*}
\begin{center}
\caption{BMW-C Parameters}
\label{bmwc:params}
\begin{tabular}{rllccccc}
\hline
\hline
\noalign{\smallskip}
Column & BMWC Parameter & Description \\
\noalign{\smallskip}
\hline
\noalign{\smallskip}
1  & UNIQ\_NUM	& source unique number (internal reference)	\\ 
2  & SRC\_NAME  & source IUA name BMCHHMMSS.S$\pm$DDMMSS 	 \\ 
3  & SEQ\_NUM 	& sequence Number of the Observation  		(e.g., 900030) \\ 
4  & OBJECT 	& target name 					(e.g., HDFNORTH)	\\ 
5  & OBS\_ID 	& observation ID Number	of the Target 		(e.g., 2386)\\  
6  & EXPOSURE 	& total exposure time 			(s)	\\  
7  & EXPTIME 	& commanded exposure time 		(s)	\\  
8  & ASCDSVER	& ASCDS version number				(e.g., 6.0.1)\\   
9  & DATE\_OBS	& date and time of observation start 		(e.g., 2000-11-20T05:39:48)\\
10 & DATE\_END  & date and time of observation stop  		(e.g., 2000-11-20T08:54:42)\\  
11 & RA\_NOM	& nominal target right ascension 		(RA; J2000, degrees) 	\\ 
12 & DEC\_NOM	& nominal target declination 			(DEC; J2000, degrees) 	\\ 
13 & OBS\_MODE 	& observation mode 				(e.g., POINTING) \\ 
14 & DATAMODE 	& data mode 					(e.g., FAINT) 	\\   
15 & READMODE 	& read mode  					(e.g., TIMED) 	\\  
16  & TEFF  		& effective exposure (s, after flares are removed)	\\ 
17  & T\_BCK   		& total-band mean background in the central $1024\times1024$ pix region (cts/pix)\\ 
18  & S\_BCK     	& soft-band mean background in the central $1024\times1024$ pix region  (cts/pix)\\ 
19  & H\_BCK  		& hard-band mean background in the central $1024\times1024$ pix region  (cts/pix)\\ 
20  & NH\_WAVG		& weighted averaged $N_{\rm H}$ (cm$^{-2}$) for RA\_NOM, DEC\_NOM \\
21  & COR\_FAC1		& (erg cm$^{-2}$ cts$^{-1}$), counts(0.5--7 keV) vs.\ Flux(0.5--10 keV) for field $N_{\rm H}$  \\
22 & SRC\_RA\_H & source RA 				(hh) \\
23 & SRC\_RA\_M & source RA 				(mm) \\
24 & SRC\_RA\_S & source RA 				(ss) \\
25 & SRC\_DE\_D & source DEC 				(dd) \\
26 & SRC\_DE\_M & source DEC 				(mm) \\
27 & SRC\_DE\_S & source DEC 				(ss) \\
28  & SRC\_ALPH	& source RA 				(J2000, degrees)\\  
29  & SRC\_DELT	& source DEC 				(J2000, degrees)\\  
30  & X\_POS\_ER& error on source RA 			(arcsec)	\\  
31  & Y\_POS\_ER& error on source DEC  			(arcsec)\\  
32  & T\_POS\_ER& total positional error 		(arcsec)\\  
33  & WIDTH  	& width of WT Gaussian distribution of photons	(arcsec)\\  
34  & WIDTH\_ER & error on width 			(arcsec)\\ 
35  & W\_FLAG  	& meaningfulness of width (0/1; 0$=$free, $1=$fixed to PSF)	\\  
36  & SCALE  	& scale	at which source was found 	(1,2,4,8,16,32,64 pixels)		\\  
37  & CHI2 	& fit $\chi^2$								\\  
38  & OFFAX  	& offaxis angle					(arcmin) 		\\  
39  & X\_POS  	& source x-coordinate				(pixel) 		\\  
40  & Y\_POS  	& source y-coordinate 				(pixel) 		\\  
41  & T\_FT\_CTS 	& total-band (0.5--7\,keV) counts from the fit (cts) \\
42  & T\_CTS\_ER 	& total-band error on counts from the fit (cts)	\\ 
43  & T\_CN\_CTS  	& total-band net counted source counts 	(cts)\\ 
44  & T\_CN\_BG  	& total-band counted background counts  (cts)	\\ 
45  & T\_VIGCOR  	& total-band vignetting correction factor 	\\ 
46  & T\_PSFCOR  	& total-band PSF correction factor 	\\ 
47  & T\_CTRATE  	& total-band count rate 		(cts s$^{-1}$)	\\ 
48  & T\_CR\_ER		& total-band count rate error 		(cts s$^{-1}$) \\
49  & T\_FLUX 		& total-band flux  (for CF0$=1\times10^{-11}$ erg cm$^{-2}$ cts$^{-1}$) \\
50  & T\_FT\_S\_N	& total-band signal-to-noise ratio of the detection \\ 
51  & T\_WV\_S\_N	& total-band signal-to-noise ratio in wavelet space \\
52  & S\_FT\_CTS 	& soft-band (0.5--2\,keV) counts from the fit (cts) \\
53  & S\_CTS\_ER 	& soft-band error on counts from the fit (cts)\\ 
54  & S\_CN\_CTS 	& soft-band net counted source counts	(cts)\\ 
55  & S\_CN\_BG  	& soft-band counted background counts	(cts)\\ 
56  & S\_VIGCOR  	& soft-band vignetting correction factor	\\ 
57  & S\_PSFCOR  	& soft-band PSF correction factor	\\ 
58  & S\_CTRATE  	& soft-band count rate    		(cts s$^{-1}$)\\
59  & S\_CR\_ER		& soft-band count rate error		(cts s$^{-1}$)	\\
60  & S\_FLUX  		& soft-band flux  (for CF0$=1\times10^{-11}$ erg cm$^{-2}$ cts$^{-1}$)	\\ 
61  & S\_FT\_S\_N  	& soft-band signal-to-noise ratio of the detection	\\ 
62  & S\_WV\_S\_N  	& soft-band signal-to-noise ratio in wavelet space	\\ 
\noalign{\smallskip}
\hline
\end{tabular}
\end{center}
\end{table*}

       \setcounter{table}{2}
\begin{table*}
\begin{center}
\caption{BMW-C Parameters (Continued)}
\begin{tabular}{rllccccc}
\hline
\hline
\noalign{\smallskip}
Column & BMWC Parameter & Description \\
\noalign{\smallskip}
\hline
\noalign{\smallskip}
63  & H\_FT\_CTS 	& hard-band (2--7\,keV) counts from the fit (cts) \\
64  & H\_CTS\_ER  	& hard-band error on counts from the fit (cts)	\\ 
65  & H\_CN\_CTS  	& hard-band net counted source counts (cts)\\ 
66  & H\_CN\_BG   	& hard-band counted background counts	(cts)\\ 
67  & H\_VIGCOR   	& hard-band vignetting correction factor	\\ 
68  & H\_PSFCOR   	& hard-band PSF correction factor 	\\ 
69  & H\_CTRATE 	& hard-band count rate 			(cts s$^{-1}$)\\ 
70  & H\_CR\_ER 	& hard-band count rate error 		(cts s$^{-1}$)\\ 
71  & H\_FLUX 		& hard-band flux (for CF0$=1\times10^{-11}$ erg cm$^{-2}$ cts$^{-1}$)	\\ 
72  & H\_FT\_S\_N 	& hard-band signal-to-noise ratio of the detection	\\ 
73  & H\_WV\_S\_N 	& hard-band signal-to-noise ratio in wavelet space	\\ 
74  & FL\_REBIN		& rebin at which source was found (e.g., 1/2)\\
75  & WTPROBAB 		& WT Probability\\
76  & FLUX1 		& absorption corrected 0.5--10 keV flux (for {\tt COR\_FAC1(NH\_WAVG)})  \\
77  & FLUX2		& observed 0.5--10 keV flux  (for CF2($N_{\rm H}=0$)$=8.22981\times10^{-12}$ erg cm$^{-2}$ cts$^{-1}$) \\
78  & FL\_CHECK		& quality flag: $= 1$ status ok; $< -1$ decreasing quality\\
79  & FL\_CHCK2		& quality flag: $-1=$ to be visually inspected, $<-1$ decreasing quality; $|$ $| > 10$ pointed source\\
80  & FL\_EXT\_Y   	& flag for extension (Point/Extended) \\
81  & CORR\_RAD 	& cross-matching radius\\
82  & FL\_MERGE 	& cross-matching: part of merged catalogue? $\ga1=$yes, $0=$no \\
83  & CORR\_RA2	        & cross-matching radius\\
84  & FL\_MERG2 	& cross-matching: part of merged catalogue? $\ga1=$yes, $0=$no \\
85  & CAT\_VERS  	& catalogue version \\
\noalign{\smallskip}
\hline
\noalign{\smallskip}
86 & ID\_FIRST 	& FIRST identification\\
87 & RA\_FIRST 	& FIRST RA 	(degrees) \\
88 & DE\_FIRST 	& FIRST DEC 	(degrees) \\
89 & LOBF\_1ST	& warning flag for side-lobe source\\
90 & PFLX\_1ST	& peak flux (mJy/bm)\\
91 & IFLX\_1ST	& integrated Flux (mJy)\\
92 & RMS\_1ST	& local noise at the source position (mJy/beam)\\
93 & MAJ\_1ST 	& deconvolved major axis (FWHM in arcsec; elliptical Gaussian model)\\
94 & MIN\_1ST 	& deconvolved min axis (FWHM in arcsec; elliptical Gaussian model)\\
95 & PA\_1ST 	& deconvolved position angle (degrees, east of north) \\
96 & fMAJ\_1ST 	& measured major axis (arcsec)\\
97 & fMIN\_1ST 	& measured minor axis(arcsec)\\
98 & fPA\_1ST 	& measured position angle (deg)\\
99 & FIRSTBMC	& angular distance between FIRST and BMW-C position (arcsec) \\
100 & FIRSTCOM & number of FIRST cross-matches: $-99=$none, 1$=$single, $>1=$number of matches \\
101 & FIRSTCOV	& BMW-C--FIRST cross-matching version \\
\noalign{\smallskip}
\hline
\noalign{\smallskip}
102 & ID\_IRAS 	& IRAS identification\\
103 & RA\_IRAS 	& IRAS RA (J2000, degrees)\\
104 & DE\_IRAS 	& IRAS DEC (J2000, degrees)\\
105 & F12\_IRAS		& 12 $\mu$m flux (mJy) \\
106 & F25\_IRAS 		& 25 $\mu$m flux (mJy) \\
107 & F60\_IRAS 		& 60 $\mu$m flux (mJy) \\
108 & F100\_IRAS 		& 100 $\mu$m flux (mJy) \\
109 & IRASBMC		& angular distance between IRAS and BMW-C position (arcsec) \\
110 & IRASCOM		& number of IRAS cross-matches: $-99=$none, $1=$single, $>1=$number of matches \\
111 & IRASCOV		& BMW-C--IRAS cross-matching version \\
\noalign{\smallskip}
\hline
\noalign{\smallskip}
112 & ID\_2MASS 	& 2MASS identification\\
113 & RA\_2MASS 	& 2MASS RA (J2000, degrees)\\
114 & DE\_2MASS 	& 2MASS DEC (J2000, degrees)\\
115 & J\_2MASS 		& 2MASS $J$ magnitude (mag)\\
116 & DJ\_2MASS 		& 2MASS error on $J$ magnitude (mag)\\
117 & H\_2MASS 		& 2MASS $H$ magnitude (mag)\\
118 & DH\_2MASS 		& 2MASS error on $H$ magnitude (mag)\\
119 & K\_2MASS 		& 2MASS $K$ magnitude (mag)\\
120 & DK\_2MASS 		& 2MASS error on $K$ magnitude (mag) \\
121 & 2MASSBMC		& angular distance between 2MASS and BMW-C position (arcsec) \\
122 & 2MASSCOM		& number of 2MASS cross-matches:$ -99=$none, 1$=$single, $>1=$number of matches \\
123 & 2MASSCOV		& BMW-C--2MASS cross-matching version \\
\noalign{\smallskip}
\hline
\end{tabular}
\end{center}
\end{table*}

       \setcounter{table}{2}
\begin{table*}
\begin{center}
\caption{BMW-C Parameters (Continued)}
\begin{tabular}{rllccccc}
\hline
\hline
\noalign{\smallskip}
Column & BMWC Parameter & Description \\
\noalign{\smallskip}
\hline
\noalign{\smallskip}
124 & ID\_GSC2	& GSC2 identification	\\
125 & RA\_GSC2	& GSC2 RA (J2000, degrees)	\\
126 & DE\_GSC2	& GSC2 DEC (J2000, degrees)	\\
127 & F\_GSC2		& GSC2 $F$ magnitude (mag) \\
128 & DF\_GSC2	& GSC2 error on $F$ magnitude (mag)	\\
129 & J\_GSC2 	& GSC2 $B_J$ magnitude (mag) \\
130 & DJ\_GSC2	& GSC2 error on $B_J$ magnitude (mag) \\
131 & V\_GSC2 	& GSC2 $V$ magnitude (mag) \\
132 & DV\_GSC2	& GSC2 error on $V$ magnitude (mag) \\ 
133 & N\_GSC2 	& GSC2 $N$ magnitude (mag) \\
134 & DN\_GSC2	& GSC2 error on $N$ magnitude (mag) \\
135 & A\_GSC2 	& GSC2 semi-major axis (pixels) \\
136 & E\_GSC2 	& GSC2 eccentricity\\
137 & PA\_GSC2	& GSC2 position angle (deg)\\
138 & C\_GSC2 	& GSC2 class: 0$=$star; 1$=$galaxy; 2$=$blend; 3$=$non-star; 4$=$unclassified; 5$=$defect \\
139 & GSC2BMC		& angular distance between GSC2 and BMW-C position (arcsec) \\
140 & GSC2COM 	& number of GSC2 cross-matches: $-99=$none, $1=$single, $>1=$number of matches \\
141 & GSC2COV	& BMW-C--CSC2 cross-matching version \\
142 &  CTS05\_1  & 0.5-1 keV counted source counts\\
143 &  CTS\_1\_2  & 1-2 keV counted source counts\\
144 &  CTS\_2\_4  & 2-4 keV counted source counts\\
145 &  CTS\_4\_7  & 4-7 keV counted source counts\\
\noalign{\smallskip}
\hline
\end{tabular}
\end{center}
\end{table*}

	\section{Source properties and sub-samples\label{bmwc:character}} 

The full catalogue includes all detections in the 136 examined fields, 
and no association was made among the sources detected more than once in 
different observations of the same portion of the sky. 
In order to account for repeated detections and to exploit 
multiple detections of the same source in variability studies,   
an estimate of the number of independent source detections is needed.  
We estimated this number by cross matching the catalogue with a given  
cross-matching radius, i.e.,  
by merging the number of sources within the cross-matching radius. 
Given the distribution of total position uncertainties 
(Fig.~\ref{bmwc:poserr}), we chose cross-matching radii of 3\arcsec and 4\farcs5 
(catalogue parameters {\tt CORR\_RAD} and {\tt CORR\_RA2}, respectively). 
We obtain 16088 independent sources in the former case, and 15497 in the latter case
(of which 12135 and 11954, respectively, are  not associated with bright and/or 
extended sources at the centre of the field).  
The catalogue parameter {\tt FL\_MERGE}  assumes the value 1 if the source is part of the 
sub-sample merged within {\tt CORR\_RAD}, 0 otherwise.  
{\tt FL\_MERG2} is the corresponding value for {\tt CORR\_RA2}.

	\subsection{The serendipitous source catalogue (BMC--SSC) \label{bmwc:ssc}} 

For cosmological studies it is particularly important to have a sample which is 
not biased toward objects selected on the basis of their properties. 
To this end, we selected a subsample of the BMW-C catalogue that contains 
16758 sources not associated with pointed targets, by  
excluding sources within a radius of 30$\arcsec$ from the target position. This subsample  
represents the  BMW-{\it Chandra} Serendipitous Source Catalogue (BMC-SSC).

	\subsection{Source extension\label{bmwc:extension}} 

	\begin{figure}
	 \resizebox{\hsize}{!}{\includegraphics{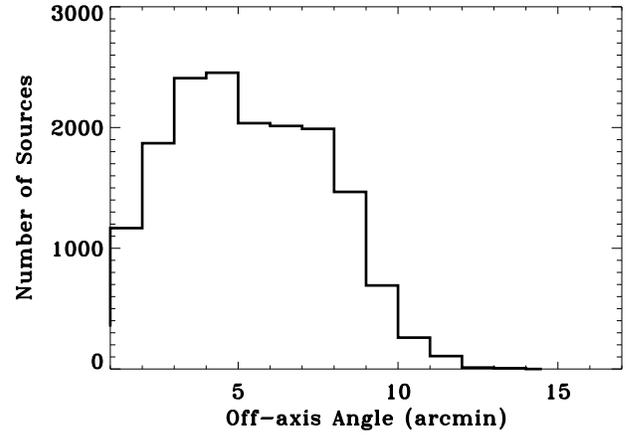}}
	 \caption{Distribution of the off-axis angle of the 16834 
		BMW-{\it Chandra} sources.}
	 \label{bmwc:offax}
	\end{figure}

Figure~\ref{bmwc:offax} shows the distribution of the source off-axis angle {\tt OFFAX}, 
which presents a steep increase with collecting area, and a gentler decrease with 
decreasing sensitivity with off-axis angles. Differently from what found with the 
BMW-HRI catalogue (\citealt{Panzeraea03}), our distribution does not present 
a peak at zero off-axis due to pointed sources. 

To characterise the source extension, 
which is one of the main features of the WT method, one cannot simply compare 
the WT width with the instrumental PSF at a given off-axis angle. 
Thus, we use a $\sigma$-clipping algorithm which divides the 
distribution of source extensions as a function of off-axis angle 
in bins of 1\arcmin{} width. 
The mean and standard deviation are calculated within each bin and all sources 
which width exceeds 3$\sigma$ the mean value are discarded. 
The procedure is repeated until convergence is 
reached. The advantage of this method is that it effectively eliminates truly 
extended sources, while providing a value for the mean and standard deviation 
in each bin (\citealt{Lazzatiea99}). The mean value plus the 3$\sigma$
dispersion\footnote{The fitting function is the third-order polynomial: 
3$\sigma$~extension~(arcsec) $=1.24097+0.0530598\,{\tt OFFAX^2} +0.000179786\,{\tt OFFAX^3}$, 
where {\tt OFFAX} is in arcmin.}
provides the line discriminating the source extension, but we conservatively classify as 
extended only the sources that lie 2$\sigma$ above this limit. 
Combining this threshold with the 3$\sigma$ on the intrinsic dispersion, 
we obtain a $\sim 4.5$$\sigma$ confidence level for the extension classification
(\citealt{Rosatiea95,Campanaea99,Panzeraea03}). 
We note (\citealt{Morettiea04}) that fluxes of extended sources are usually 
underestimated, since they are computed by fitting a Gaussian to the surface 
brightness profile, which in many cases is a poor approximation.

	\begin{figure}
	 \resizebox{\hsize}{!}{\includegraphics{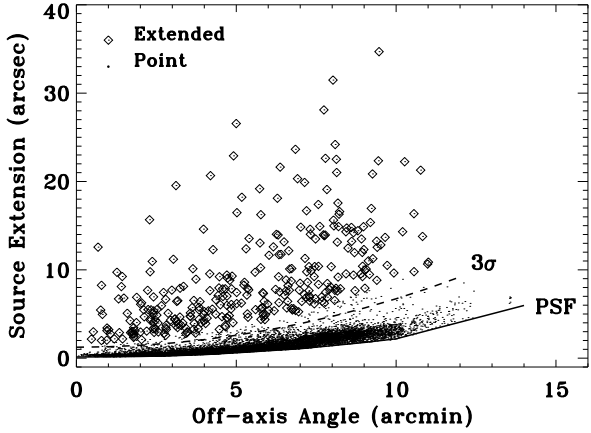}}
	 \caption{Extension of the 12648 BMW-{\it Chandra} sources as a 
	function of the off-axis angle. 
	The dashed line is the 3$\sigma$ limit for point sources. 
	Diamonds are the extended sources ($\sim 4.5 \sigma$), i.e., sources that 
	lie more than 2 $\sigma$ from the dashed line (316 points). }
		 \label{bmwc:extension_final1}
	\end{figure}

	\begin{figure}[t]
	 \resizebox{\hsize}{!}{\includegraphics{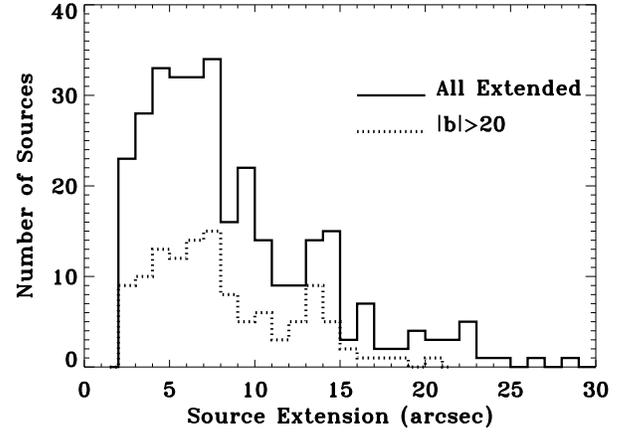}}
	 \caption{Distribution of the extension of the extended sources for the full sample
		(solid line, 316 sources) and for the high-latitude sub-sample 
		(dotted line, 120 sources).}
	 \label{bmwc:fullg_hwidthb}
	\end{figure}
We applied the $\sigma$-clipping algorithm to the 12648 good sources 
which WT width had been successfully determined 
(i.e., the width had not been fixed to the PSF value, {\tt W\_FLAG}$ = 1$). 
Figure~\ref{bmwc:extension_final1} shows their extension 
thus calculated as a function of the off-axis angle,
as well as the PSF function and 3$\sigma$ limit for point sources (dashed line).
Diamonds are the extended sources ($\sim 4.5 \sigma$), i.e., sources that 
lie more than 2 $\sigma$ from the dashed line (316 points).
There are 145 points within the 3 and 4.5$\sigma$ limits. 

We selected two sub-catalogues, based on Galactic latitude, the discriminant 
value being 20$^{\circ}$, thus obtaining 7401 high-latitude sources, 
and 9433 low-latitude sources.
Figure~\ref{bmwc:fullg_hwidthb} shows the distribution of the extension of 
the extended sources for the full sample (solid line, 316 sources) 
and for the high-latitude sub-sample (dotted line, 120 sources).
We expect that most high-galactic latitude sources are extra-galactic. 
Table~\ref{bmwc:numeri} summarises the basic numbers of sources in each subsample 
examined. 
%
%
%
%
\begin{table}
\begin{center}
\caption{BMW-C Basic Numbers.}
\label{bmwc:numeri}
\begin{tabular}{llr}
\hline
\hline
\noalign{\smallskip}
Source Sample & & 				Number \\
\noalign{\smallskip}
\hline
\noalign{\smallskip}
  detected    			&    				&    	21325  \\
  good$^{\mathrm{a}}$ 		&    				&    	16834  \\
  serendipitous 	 	&				& 	16758  \\
  independent 			& ({\tt CORR\_RAD = 3\arcsec})	&  	12135 \\
				& ({\tt CORR\_RA2 = 4\farcs5})  & 	11954 \\
  detected in total band	& 				& 	11124\\
  detected in soft band		&				& 	12631 \\
  detected in hard band		&				& 	9775\\
  only detected in hard band	&				& 	4203 \\
  high-latitude			& ($| b|>20^{\circ}$)  		&       7401 \\
                        	& not pointed  			&       7381 \\
                        	& not in surveys  		&       2931 \\
                        	& not in surveys or pointed   	&       2911 \\
  low-latitude	 		& ($| b|<20^{\circ}$)   		&    	9433 \\
                        	&  not pointed     		&       9377 \\
                        	&  not in surveys  		&     	9433 \\
                        	&  not in surveys or pointed    &   	9377 \\
  serendipitous extended 	&    				&    	316  \\
	       		  	&   ($| b|>20^{\circ}$)		&     120 \\
        		  	&   ($| b|<20^{\circ}$)		&     196 \\
\noalign{\smallskip}
\hline
\end{tabular}
\end{center}
\begin{list}{}{}
\item[$^{\mathrm{a}}$] Sources which do not require a more in-depth, non-automated analysis 
(i.e.\ not associated with bright and/or extended sources at the centre of the field), 
including the target ones.
\end{list}
\end{table}

	\subsection{Source fluxes\label{bmwc:fluxes}} 

\begin{figure}[t]
         \resizebox{\hsize}{!}{\includegraphics{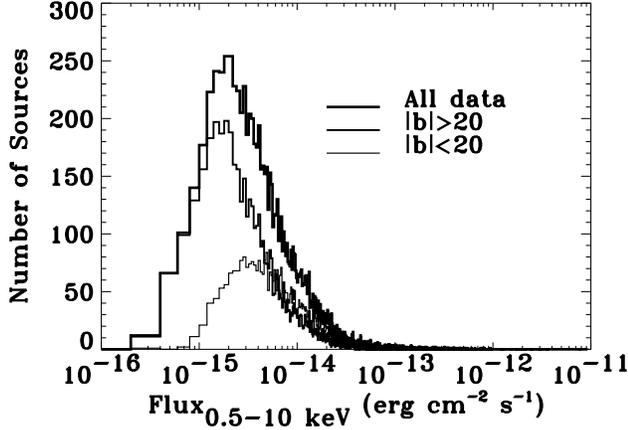}}
         \caption{Distributions of 0.5--10\,keV flux for 
		the sources in the BMW-{\it Chandra} catalogue
		({\tt FLUX1}, the Galactic absorption corrected flux), 
		for the high-latitude sources, 
		and the low-latitude sources.}
         \label{bmwc:histo_flux}
        \end{figure}

\begin{figure}
         \resizebox{\hsize}{!}{\includegraphics{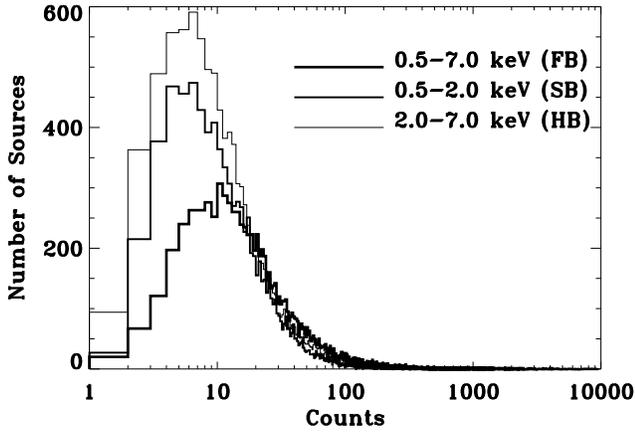}}
         \caption{Distributions of the source counts for 
	the soft, hard, and full bands, drawn from the full catalogue.}
         \label{bmwc:histo_ctcnt}
        \end{figure}

Figure~\ref{bmwc:histo_flux} shows the distributions of 0.5--10\,keV absorption 
corrected flux ({\tt FLUX1}, see Sect.~\ref{bmwc:catalogue}) for the 16834 sources in the 
BMW-{\it Chandra} catalogue; 
the fluxes range from $\sim 3\times 10^{-16}$ to $9\times10^{-12}$
erg cm$^{-2}$ s$^{-1}$ with a median of $7\times 10^{-15}$ erg cm$^{-2}$ s$^{-1}$.
Figure~\ref{bmwc:histo_flux} also shows the distributions of 0.5--10\,keV fluxes 
for the high-latitude sources (median flux $4.50\times 10^{-15}$ erg cm$^{-2}$ s$^{-1}$)   
and the low-latitude sources  (median flux $1.07\times 10^{-14}$ erg cm$^{-2}$ s$^{-1}$). 
Figure~\ref{bmwc:histo_ctcnt} shows the distributions of the source counts for 
	the total (FB, median   27.9  counts), 
	soft (SB, median 15.7   counts), 
	and hard  (HB, median   15.3 counts) bands. 
	%

	\section{Sky coverage \label{bmwc:skycov_how}} 

In order to calculate the sky coverage of our survey, we 
followed the procedure used by \citet{Munoea03}. 
The signal-to-noise ratio S/N with which we measure the counts 
from a source is given by 
$n_{\sigma} = N/[(N + B) + \sigma_B^2]^{1/2}$, 
where $N$ is the number of counts from a source, $B$ is the background 
in the source cell and $\sigma_B$ is the uncertainty on the background, 
using the simplifying assumption of $\sqrt{N}$ uncertainties. 
The background can be written as $B = ab$, where $a$ is the area of 
the PSF and $b$ is the background per pixel. Therefore, using 
$\sigma_B^2 = B$, the  S/N  becomes   
$n_{\sigma} = N/(N + 2B)^{1/2} = N/(N + 2ab)^{1/2} $. 
The position-dependent number of source counts for a given S/N is then, 
\begin{equation}
S = {{n_{\sigma}^2} \over {2}}  
    \left[ 1 + \left(1 + {{8ab}\over{n_{\sigma}^2}} \right)^{1/2} \right]
		 {\rm [cts]}
\label{bmwc:cslimeq}
\end{equation}
First, an off-axis angle map is generated, then 
the PSF maps $a$ are generated from the off-axis map using the 
psfsize table in CALDB, 
using the PSF images at $\sim 1.50$\,keV for the soft band PSF 
                and at $\sim 4.51$\,keV for the hard band PSF.  
We assumed as aperture the radius that encircles 70\% of the PSF energy 
(as a reasonable compromise between having too many background counts 
for a larger radius and too little source counts for a smaller radius). 
For $b$ we used the background images generated for the detection 
(see Sect.~\ref{bmwc:bgdmaps}). 
        \begin{figure}
         \resizebox{\hsize}{!}{\includegraphics{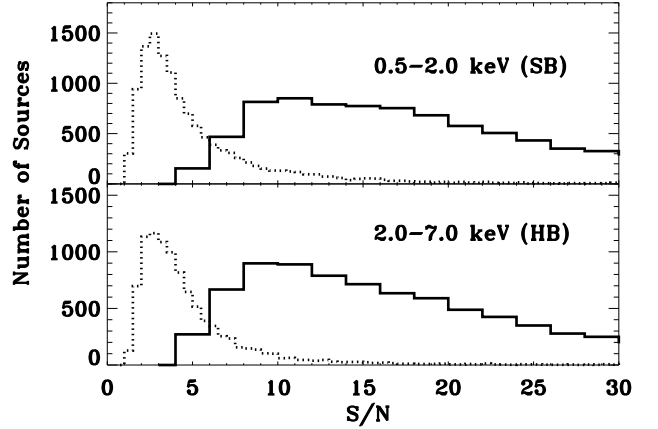}}
         \caption{S/N distribution in the counts space (dotted line) 
		and wavelet space (solid line) in the 
		soft ({\bf top}) and hard ({\bf bottom}) bands.}
         \label{bmwc:histo_sn}
        \end{figure}
	
        \begin{figure} 
         \resizebox{\hsize}{!}{\includegraphics{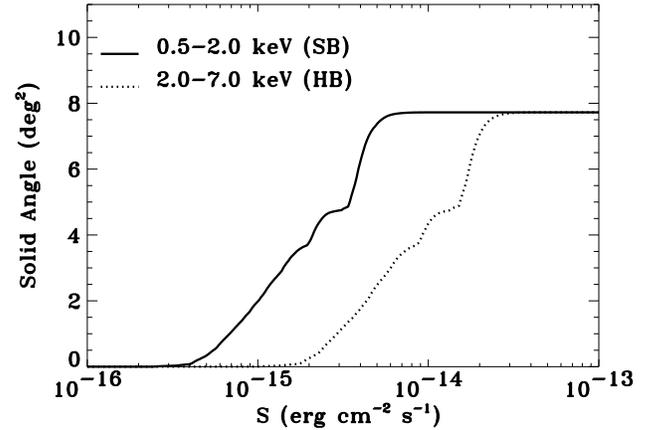}}
         \caption{Solid angle versus flux limit for S/N $=3$ for the 
	soft (solid line) and hard (dotted line) bands, 
	calculated following Sect.~\ref{bmwc:skycov_how}. This sky coverage 
	was constructed using 94 independent fields (no fields covered the same sky area).}
         \label{bmwc:sky_both_sn3}
        \end{figure}

Based on Fig.~\ref{bmwc:cha_prob}, which shows the wavelet probability as a 
function of WT S/N, and background value, we chose S/N$=3$ in the wavelet space, 
for our analysis.
Figure~\ref{bmwc:histo_sn} shows the S/N distribution in the counts space and wavelet space in 
the soft and hard bands, and demonstrates where our S/N cut falls.
We note here that the S/N in WT space can be different
from the S/N in counts space for several reasons: 
i) the background subtraction is more accurate and locally performed; 
ii) the high frequencies are suppressed, so that a correlated count 
excess gives a higher significance than a random one with the same number of counts;
iii) the exposure map can be incorporated in the WT space, so
that artifacts do not affect the source significance.
Using Eq.~\ref{bmwc:cslimeq}, we calculated the limiting counts for each 
field, in the soft and hard bands. 
We then converted the limiting counts maps into limiting flux maps, 
using count rate-to-flux conversion factors derived assuming a power-law model 
with a photon index  $\Gamma =$1.7 in the soft band and  $\Gamma =$1.4 in the hard band,
modified by the absorption by Galactic $N_{\rm H}$ relative to each field. 
The adopted values of $\Gamma$ were chosen to compare with the results in the literature
(e.g.\ \citealt{Champ3,Champ2}).
Histograms of the number of pixels (hence solid angle) 
with flux smaller than a given threshold were produced for each field 
in the soft and hard bands for S/N$=3$, and the solid angle 
of the whole survey calculated as the sum of the contributions of each field.
We note here that some of the observations covered the same sky area, notably so for the 
survey fields. Therefore, only the observation with the longest exposure time (after 
screening and correction of the data) was considered for the sky coverage 
calculation.
In Fig.~\ref{bmwc:sky_both_sn3}  we show the 
solid angle of the full survey (94 independent fields) as a function of the 
flux limit for S/N$=$3.
The complete catalogue  provides a sky coverage in the soft band (0.5--2\,keV, S/N =3)
of $\sim 8$ deg$^2$ at a limiting flux of $\sim 10^{-13}$  erg cm$^{-2}$ s$^{-1}$,
and $\sim 2$ deg$^2$ at a limiting flux of $\sim 10^{-15}$ erg cm$^{-2}$ s$^{-1}$ 
(in the soft band).

		\section{Comparison with the ChaMP catalogue\label{bmwc:champcat}}

The {\it Chandra} Multiwavelength Project (ChaMP) is a survey of serendipitous {\it Chandra} 
sources carried out by a multi-institution 
collaboration\footnote{http://hea-www.harvard.edu/CHAMP/.}. 
It covers $\sim 10$ deg$^2$ at flux levels of 
$\sim 9\times 10^{-16}$ erg cm$^{-2}$ s$^{-1}$ (\citealt{Champ1,Champ2,Champ3}), 
and is derived from the analysis of 149 {\it Chandra} fields, that include both 
ACIS-I and ACIS-S images and a total of $\sim 6\,800$ sources. 

\begin{figure}
	 \resizebox{\hsize}{!}{\includegraphics{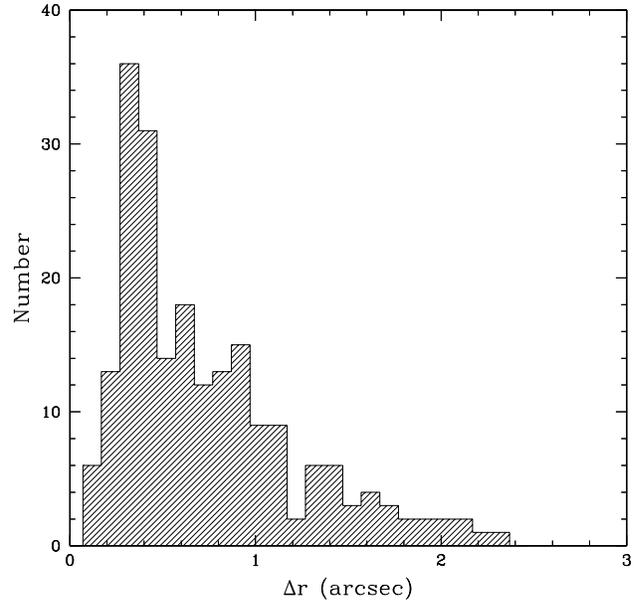}}
	 \caption[BMW-C--ChaMP: angular separation]{BMW-C--ChaMP: 
		distribution of the angular separation between the  
		ChaMP and the BMW-C X-ray positions (210 matches). }
	 \label{bmwc:champ_distance}
	\end{figure}

\begin{figure}
	 \resizebox{\hsize}{!}{\includegraphics{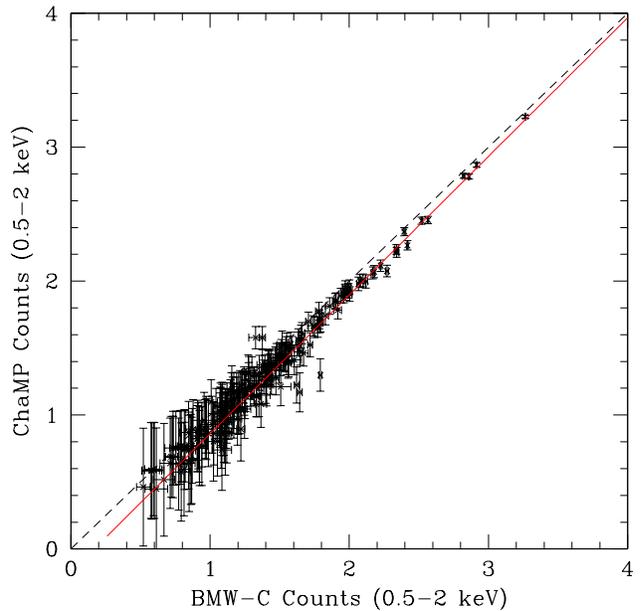}}
	 \caption[BMW-C--ChaMP: fluxes]{BMW-C--ChaMP: the 0.5--2\,keV BMW-C counts versus the 
	0.5--2\,keV ChaMP counts (210 matches). The solid line is the best fit, while 
	the dashed line is the bisecant of the plane.}
	 \label{bmwc:champ_fluxes}
	\end{figure}

The cross-matching of the source coordinates
yields 162 matches within 1\arcsec{} and 210 within 3\arcsec; 
shifting the coordinates of our source list by 1\,\arcmin{} and then 
cross-matching with the ChaMP again, we found no mismatches 
(a null misidentification probability). 
Fig.~\ref{bmwc:champ_distance}, shows the distribution of the angular 
separation between the  BMW-C and ChaMP positions (\citealt{Champ3}, Table~5) for the matching objects. 
In Fig.~\ref{bmwc:champ_fluxes} we show the  BMW-C  0.5--2.0\,keV counts 
versus the ChaMP  0.5--2.0\,keV counts, as well as the best fit (solid line) 
and the bisecant of the plane (dashed line),   
$Counts_{\rm \,\,\,ChaMP}= (1.035\pm0.002)+(-0.174\pm0.005)\, Counts_{\rm \,\,\,BMW-C}$. 
Although our counts are generally higher (at the percent level) 
than the ones reported by ChaMP, we note a general consistence 
within 1-$\sigma$. We also note that in general the WT algorithm is 
better equipped at disentangling sources in crowded fields and at low S/N.

	\section{Cross-match with existing databases \label{bmwc:xcorr}} 

\begin{table*}
\begin{center}
\caption{Catalogue Data for Cross-matching.}
\label{bmwc:allcats}
\begin{tabular}{lllrl}
\hline
\hline
\noalign{\smallskip}
Catalogues & Wavelength	& Sky fraction & Entries      & References\\
           & 	        & (\%)         &              &  \\
\noalign{\smallskip}
\hline
\noalign{\smallskip}
FIRST	& 20 cm			  & $\sim 25$		& 811\,117 		&	\citet{FIRST97} \\
IRASPSC	& 12, 25, 60, 100 $\mu$m  & $\sim 98$		& $\sim $250\,000 	&	\citet{IRAS88}  \\
2MASS	& 1.25, 1.65, 2.16 $\mu$m & all-sky		& 470\,992\,970		&	\citet{2MASS}  \\
GSC2	& $B_J, F, N, V$	  & all-sky ($B_J, F, N$)  & $\sim  10^9$  	&	\citet{GSC2}  \\	
\noalign{\smallskip}
\hline
\end{tabular}
\end{center}
\end{table*}

\begin{table*}
\begin{center}
\caption{ Cross-matches with other catalogues.}
\label{bmwc:allxcorrs}
\begin{tabular}{lrrrrrrr}
\hline
\hline
\noalign{\smallskip}
Catalogues &  Cross-matching  & Total  & Single   & Double & Multiple & Closest
& Expected  (\%)     \\
           &  Radius (\arcsec)  & Matches       & Matches &   Matches        & Matches     & Matches 
       & Mismatches     \\
\noalign{\smallskip}
\hline
\noalign{\smallskip}
FIRST	& 10 			& 50 		& 3	& 12 	&   35 	& 13     & 46 		\\
IRASPSC	& 20  			& 156 		&51 	& 38 	&   67 	& 87     & 49  	 \\
2MASS	& 4.5 			& 7\,687 	&5881 	& 1486	&  320 	& 6700 	 & 50 	\\
GSC2	& 4.5 			& 5\,854 	&3788  	& 962 	& 1104 	& 4485 	 & 33 	\\	
\noalign{\smallskip}
\hline
\end{tabular}
\end{center}
\end{table*}

We cross-matched our catalogue with some of the largest catalogues available at 
other wavelengths, from radio to optical, namely, 
the Faint Images of the Radio Sky at Twenty-cm (FIRST), 
the Infrared Astronomical Satellite (IRAS) Point Source catalogue (PSC), 
the Two Micron All Sky Survey (2MASS), and 
the Guide Star Catalogue 2 (GSC2). 
Table~\ref{bmwc:allcats} summarises the main characteristics of the catalogues we considered. 
In this section we report the results of the cross-match based on the 
closest-match criterion. 

The FIRST is a survey that  
covers $\sim 10\,000$ square degrees of the North and South Galactic Cap,  
which produced images with 1$\farcs$8 pixels, a typical rms of 0.15 mJy, 
a sensitivity of $\sim 1$ mJy,  and a resolution of 5\arcsec{} (\citealt{FIRST95}).
The 2003 release\footnote{http://sundog.stsci.edu/first/catalogs/readme\_03apr11.html.}
contains $\sim 800\,000$ sources and covers a total of $\sim 9\,000$ 
square degrees (\citealt{FIRST97}). 

IRAS conducted an all-sky survey at
12, 25, 60, and 100\,$\mu$m that led to the IRAS Point Source catalogue (PSC).
The PSC contains $\sim 250\,000$ sources (\citealt{IRAS88}), and away from confused 
regions of the sky, is complete to $\sim0.4$, 0.5, 0.6, and  1.0 Jy at 12, 25, 60, 
and 100\,$\mu$m, respectively, with angular resolution of $\sim 0\farcs5$ at 12\,$\mu$m  
and $\sim 2$\arcmin{}  at 100\,$\mu$m.
Typical position uncertainties are  $\sim2$ to 6\arcsec{} in-scan and about 
$\sim8$ to 16\arcsec{}  cross-scan.

The 2MASS (\citealt{2MASS}) covers virtually all 
sky with simultaneous observations in 
$J$ (1.25\,$\mu$m), $H$ (1.65\,$\mu$m), and $K_{\rm s}$ (2.17 $\mu$m) bands
with nominal magnitude limits of 15.8, 15.1, and 14.3 mag, 
for point sources, and 15.0, 14.3, and 13.5 mag for extended sources, 
respectively.
The All-Sky Data 
Release\footnote{http://www.ipac.caltech.edu/2mass/releases/allsky/index.html.}
contains positional and photometric information for $\sim 470$ million 
point sources and $\sim 1$ million extended sources. 

In the optical, we considered the 
GSC2 (\citealt{GSC2}),  
which is an all-sky catalogue of approximately 1 billion stars and galaxies. 
In particular, we used the 
last version (GSC2.3), which covers the entire sky in the $B_J$,  $F$ and $N$ bands
(roughly comparable to Johnson $B$, $R$ and $I$ bands), down to the limiting magnitudes
22.5--23, 20--22, and 19.5.
Furthermore, partial sky coverage is available in the $V$ band 
(similar to Johnson $V$ filter) down to 19.5 magnitudes. 
The astrometry, which is calibrated with the Hipparcos (\citealt{Hipparcos97})
and the Tycho 2 (\citealt{Tycho2000}) catalogues is accurate to within 0\farcs3. 
The GSC2 also provides morphological classification for objects observed at least 
in two bands with a $\sim 90 \%$ confidence level for objects at 
$\mid b \mid \ga 5^{\circ}$ and brighter than $B_{J} \sim 19$ mag.

\begin{figure}
	 \resizebox{\hsize}{!}{\includegraphics{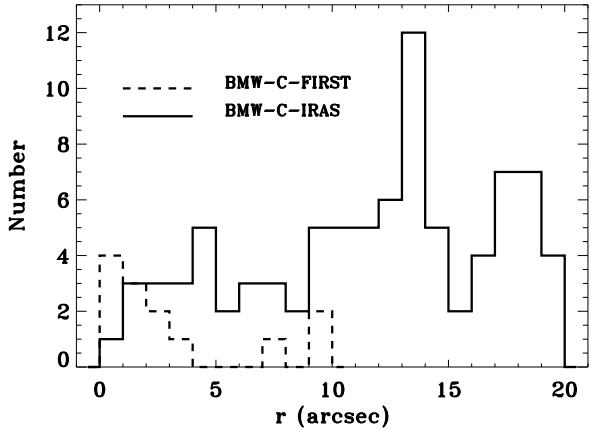}}
	 \caption{Distribution of the angular separation:  
		BMW-C--FIRST (dashed line, 13 unique matches), 
		and BMW-C--IRAS (solid line, 87 matches).}
	 \label{bmwc:distance1}
	\end{figure}

\begin{figure}
	 \resizebox{\hsize}{!}{\includegraphics{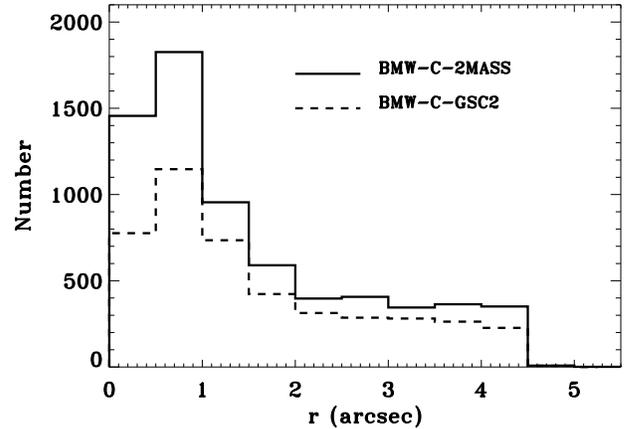}}
	 \caption{Distribution of the angular separation: 
		BMW-C--2MASS (solid line, 6700 matches) and 
		BMW-C--GSC2 (dashed line, 4458 matches).}
	 \label{bmwc:distance2}
	\end{figure}

The FIRST, IRAS, 2MASS, and GSC2 parameters are listed in 
Table~\ref{bmwc:params}.
Our cross-match procedure consisted in matching objects in the 
BMW-C catalogue with objects in the other catalogues within a given 
cross-matching radius (Table~\ref{bmwc:allxcorrs}), and 
in case of multiple matches selecting the spatially closest one.
Our choice of cross-matching depended on a combination of 
the positioning accuracy of the BMW-C and the other catalogues. 
In general, more than one optical identification was available for most 
BMW-C objects. 
Table~\ref{bmwc:allxcorrs} shows the number of BMW-C sources that have 
only one, only two, or three or more matches in the other catalogues 
(Columns 4, 5, and 6, respectively). 
The table also shows the total number of matches obtained after a 
closest distance selection (Column 7).
In Figs.~\ref{bmwc:distance1} and \ref{bmwc:distance2} we show the distribution 
of the angular separations between the BMW-C position and the position of the other 
catalogues for these closest-matching objects. For the 2MASS and GSC2, the 
distance distribution peaks at $\sim 1\arcsec$.

There is only one BMW-C source
which is found in the FIRST, IRAS, 2MASS and GSC2 catalogues (1BMC163608.2+410509); 
it is a galaxy 
observed by Infrared Space Observatory (ISO), in the European Large Area ISO Survey 
(ELAIS, \citealt{Vaisanenea02}) and in the radio (\citealt{Ciliegiea99}).
Furthermore, 71 BMW-C sources have IRAS, 2MASS and GSC2 matches; 
4208 are found in both 2MASS and GSC2 catalogues.

We calculated the expected number of mismatches (Column 8 in Table~\ref{bmwc:allxcorrs}) 
by shifting the coordinates of our source list by 1\,\arcmin{} and then cross-matching 
with the other catalogues again.

	\section{Summary and future work\label{bmwc:summary}} 

We presented the BMW-C source catalogue drawn from 136 
{\it Chandra} ACIS-I pointed observations with an exposure time in excess 
of 10\,ks public as of March 2003. 
The full catalogue comprises 21325 sources, 
16834 of which do not require a more in-depth, non-automated analysis 
(i.e.\ not associated with bright and/or extended sources), 
including the pointed ones.
Among them, 16758 are serendipitous. 
This makes our catalogue the largest compilation of {\it Chandra} sources to date. 
The 0.5--10\,keV absorption corrected fluxes of these sources 
range from $\sim 3\times 10^{-16}$ to $9\times10^{-12}$
erg cm$^{-2}$ s$^{-1}$ with a median of $7\times 10^{-15}$ 
erg cm$^{-2}$ s$^{-1}$.

The catalogue includes count rates and relative errors in three energy bands 
(total, 0.5--7\,keV; soft, 0.5--2\,keV; and hard band, 2--7\,keV; 
the source positions are relative to the highest signal-to-noise 
detection among the three bands), 
as well as source counts extracted in four additional energy bands, 
SB1 (0.5--1\,keV), SB2 (1.0--2.0\,keV), HB1 (2.0--4.0\,keV) and HB2 (4.0--7.0\,keV), 
and information drawn from the headers of the original files.
The WT algorithm also provides an estimate of the extension of 
the source which we refined with a $\sigma$-clipping method.

We computed the sky coverage for the full catalogue and for a subset
at high Galactic latitude ($\mid b \mid \, > 20^{\circ}$).
The complete catalogue provides a sky coverage in the soft band (0.5--2\,keV, S/N =3)
of $\sim 8$ deg$^2$ at a limiting flux of $\sim 10^{-13}$  erg cm$^{-2}$ s$^{-1}$,
and $\sim 2$ deg$^2$ at a limiting flux of $\sim 10^{-15}$ erg cm$^{-2}$ s$^{-1}$.

We also presented the results of the cross-match with existing catalogues
at different wavelengths (FIRST, IRAS, 2MASS, GSC2 and ChaMP).

Among the scientific applications of the catalogue are: 
1) search for periodic and non-periodic 
variability using light curves extracted for bright sources;  
2) the optical/IR follow-up of a list of galaxy cluster candidates drawn from 
our sub-sample of $\sim 300$ extended sources; 
3) analysis of blank fields, i.e.\ X-ray detected sources without counterparts 
at other wavelengths; 
4) optical/IR follow-up of peculiar sources, such as 
 isolated neutron stars candidates (ultra-soft sources) and  heavily absorbed sources 
(ultra-hard sources, not observed in the soft bands).

The current version of the BMW-C source catalogue,
(as well as additional information and data) is available at the
Brera Observatory site: {\tt http://www.brera.inaf.it/BMC/bmc\_home.html}.

\begin{acknowledgements}

We thank the anonymous referee for insightful comments that improved our paper. 
This work was supported through Consorzio Nazionale per l'Astronomia
e l'Astrofisica (CNAA) and Ministero dell'Istruzione, 
dell'Universit\`a{} e della Ricerca (MIUR) grants. 
We thank A.\ Mist\`o{} for his help with the database software. 
This publication makes use of data products from the {\it Chandra} Data Archive, 
the FIRST, IRAS, 2MASS, GSC2 surveys. 
\end{acknowledgements}

\end{document}